\documentclass[twocolumn]{aastex631}

\usepackage{physics}
\newcommand{\msun}{{\rm M}_\odot}

\begin{document}

\title{Tracking Star-Forming Cores as Mass Reservoirs in Clustered and Isolated Regions Using Numerical Passive Tracer Particles}

\correspondingauthor{Shingo Nozaki}
\email{nozaki.shingo.307@s.kyushu-u.ac.jp}

\author[0000-0003-4271-4901]{Shingo Nozaki}
\affiliation{Department of Earth and Planetary Sciences, Faculty of Science, Kyushu University, Nishi-ku, Fukuoka 819-0395, Japan}

\author[0000-0002-0547-3208]{Hajime Fukushima}
\affiliation{Center for Computational Sciences, University of Tsukuba, Ten-nodai, 1-1-1 Tsukuba, Ibaraki 305-8577, Japan}

\author[0000-0002-2062-1600]{Kazuki Tokuda}
\affiliation{Department of Earth and Planetary Sciences, Faculty of Science, Kyushu University, Nishi-ku, Fukuoka 819-0395, Japan}
\affiliation{National Astronomical Observatory of Japan, National Institutes of Natural Science, 2-21-1 Osawa, Mitaka, Tokyo 181-8588, Japan}

\author[0000-0002-0963-0872]{Masahiro N. Machida}
\affiliation{Department of Earth and Planetary Sciences, Faculty of Science, Kyushu University, Nishi-ku, Fukuoka 819-0395, Japan}

\begin{abstract}
Understanding the physical properties of star-forming cores as mass reservoirs for protostars, and the impact of turbulence, is crucial in star formation studies. We implemented passive tracer particles in clump-scale numerical simulations with turbulence strengths of $\mathcal{M}_{\rm rms} = 2, 10$. Unlike core identification methods used in observational studies, we identified 260 star-forming cores using a new method based on tracer particles falling onto protostars. Our findings reveal that star-forming cores do not necessarily coincide with high-density regions when nearby stars are present, as gas selectively accretes onto protostars, leading to clumpy, fragmented structures. We calculated convex hull cores from star-forming cores and defined their filling factors. Regardless of turbulence strength, convex hull cores with lower filling factors tend to contain more protostars and have larger masses and sizes, indicating that cores in clustered regions are more massive and larger than those in isolated regions. Thus, the filling factor serves as a key indicator for distinguishing between isolated and clustered star-forming regions and may provide insights into the star formation processes within clustered regions. We also found that most convex hull cores are gravitationally bound. However, in the $\mathcal{M}_{\rm rms} = 10$ model, there are more low-mass, unbound convex hull cores compared to the $\mathcal{M}_{\rm rms} = 2$ model. In the $\mathcal{M}_{\rm rms} = 10$ model, 16\% of the convex hull cores are unbound, which may be explained by the inertial-inflow model. These findings highlight the influence of turbulence strength on the mass and gravitational stability of cores. 
\end{abstract}


\section{Introduction} \label{sec:intro}
The stellar initial mass function (IMF), proposed by \citet{salpeter1955}, is a crucial concept in astrophysics. Understanding what determine the IMF can lead to unravelling the mysteries of stellar evolution and the matter cycle within galaxies. Past studies have shown that IMF seems to be universal not only in nearby star-forming regions but also in extragalactic star-forming regions \citep{kroupa2001, bastian2010}. Recent observations have revealed that the power-law slope of the high-mass stars in the IMF is consistent with pioneering studies \citep[e.g., ][]{salpeter1955,kroupa2001,chabrier2003}, as discussed in \cite{Kirkpatrick2024}.
While observations clarifies the universality of the IMF, understanding the origin of the IMF is still an open question.

As a clue to investigating the origin of the IMF, the Core Mass Function (CMF) of molecular cloud cores, which are the precursors of stars, has been receiving considerable attention. \cite{motte1998} identified 60 small-scale clumps in $\rho$ Ophiuchi central region and indicated that the mass distribution of these clumps is similar to the distribution of the IMF. The link between the IMF and the CMF has also been suggested in other star-forming regions, such as Taurus and Orion \citep[e.g., ][]{tatematsu1993, onishi2002, nutter2007}. To statistically estimate the CMF, recent observations have identified many molecular cloud cores, including starless, prestellar, and protostellar cores, using core identification algorithms \cite[e.g., ][]{andre2010, konyves2015, konyves2020, marsh2016, chen2019, ladjelate2020, takemura2021a, takemura2021b}. These studies have strengthened the connection between the IMF and the CMF.

Although the observational studies described above have been very effective in revealing the general properties of dense cores, it should be noted that the dense cores identified by observations do not necessarily determine the mass of the main sequence stars. The internal structure of molecular clouds dominated by supersonic turbulence is complex, and it cannot be ruled out that components not directly related to the targeted dense cores are mixed in along the line of sight \citep[e.g., ][]{williams1994,goodman2009}.

In addition to the complexities arising from these observational limitations, the dense cores in cluster-forming regions have inherently more complex distributions than those of isolated star-forming cores. Because a large number of prestellar and protostellar cores are densely packed within 0.1-1 pc \citep[e.g., ][]{ward-thompson2007, tokuda2022} in cluster-forming regions, it becomes even more difficult to precisely define the mass supply regions for individual stars. Many small-scale ($\leq 0.1 \, \rm pc$) substructures within molecular clouds that are not gravitationally bound have been discovered \citep[e.g., ][]{chen2019, takemura2021a, sato2023}, leaving room for doubt as to whether they will eventually evolve into individual stars. There is therefore a strong demand for theoretical studies that can track dense cores as mass reservoirs for protostars.

Theoretically, the relationship between prestellar cores and protostars has been investigated through numerical simulations at both the clump scale and isolated core scales. Using numerical simulations that assume isolated dense cores as initial conditions, \citet{machida2012} showed that the ratio of final stellar mass to the mass of the isolated prestellar core is 30-60\%. Considering an isolated dense core and adopting external density distributions as a parameter, \citet{nozaki2023} suggested that the external high-density affects the final stellar mass and star formation efficiency. Observational studies also suggest the presence of mass inflow from outside the prestellar core \citep{Redaelli2022}. 
Thus, prestellar cores with higher surrounding densities are influenced by external mass accretion. 

The relationship between prestellar cores and protostars has been statistically investigated in clump-scale numerical simulations \citep[e.g., ][]{padoan2007, smith2009, haugbolle2018, padoan2020, smullen2020, pelkonen2021, offner2022, david2023, david2024}. \citet{pelkonen2021} performed high-resolution clump-scale numerical simulations to investigate the correlation between progenitor cores and final stellar mass, and identified the progenitor cores of each protostar using the clumpfind algorithm \citet{padoan2007}. Although they showed that the peak of the CMF is close to that of the IMF, they did not explicitly present a direct correlation between progenitor core masses and final stellar masses. 
Their study further revealed that the significant fraction of the mass reservoir of stars are located outside the progenitor cores regardless of stellar mass, in which the gas motion was traced by passive particles. 

Understanding the physical properties of star-forming core regions as the mass reservoir for protostars is crucial for determining the final stellar mass, which should be realized through numerical simulations using passive trace particles, as in \citet{pelkonen2021}. Turbulence could play a crucial role in the star formation process, influencing the IMF and fragmentation of molecular clouds \citep[e.g., ][]{padoan2002, maclow2004}. Observationally, parsec-scale molecular clouds can exhibit velocity dispersions ranging from sonic to supersonic speeds \citep[e.g., ][]{Falgarone1992, Rathborne2007, Battersby2010}. However, the physical properties of the mass reservoir regions for protostars and the effect of turbulence are still not fully understood.

This paper aims to investigate the physical properties of star-forming cores as the mass reservoir for each protostar, and the impact of turbulence on these properties. We conduct clump-scale numerical simulations with turbulence strengths as a parameter. Unlike the core identification algorithms used in observational studies, we identified and analyzed the star-forming cores as the mass reservoir for each protostar. Our new analysis method constructed using passive tracer particles, allows us to accurately determine the mass reservoir for each protostar. We also discuss the differences in the physical properties of star-forming cores that form in different turbulence environments.

This paper is structured as follows. Section \ref{sec:methods} describes the numerical settings and the analysis methods to identify star-forming cores. In Section \ref{sec:mass_reservoir}, the shape of the star-forming cores as the mass reservoir are presented. Section \ref{sec:convex} describes the correlation between the filling factor and the physical properties of the star-forming cores. In section \ref{sec:discussion}, we discuss the impact of turbulence strength on star-forming cores and compare these with observations of dense cores. Section \ref{sec:summary} summarizes our conclusions.

\section{Methods} \label{sec:methods}
\subsection{Numerical Settings} \label{subsec:settings}
We have conducted three-dimensional hydrodynamic simulations with SFUMATO, an adaptive mesh refinement code \citep{matsumoto2007, matsumoto2015, fukushima2021}. The calculations have been performed within a cube simulation box, each side of which spans $4 \, \rm pc$, with the total mass of the box set to $3000 \, \msun$ and an initial uniform proton number density corresponding to $1365\, \rm cm^{-3}$. Periodic boundary conditions are applied to sink particles, tracer particles, and fluid dynamics. The cell width at the finest resolution level ($l_{\rm max} = 6$) is $\Delta x = 195 \, \rm au\ (=0.000977\,pc)$, ensuring resolution exceeds five times the Jeans length. Following the approach of \cite{matsumoto2015}, we treat the root mean square (rms) Mach number $\mathcal{M}_{\rm rms}$, which characterizes the initial velocity field, as a parameter, with values set to $\mathcal{M}_{\rm rms} = 2, 10$. The turbulent Mach number $\mathcal{M}_{\rm rms}$ is given by:
\begin{equation}
\mathcal{M}_{\rm rms} = \left[ \frac{1}{c_{\rm s}^2 L^3} \int_V |\mathbf{v}_t|^2 \, dV \right]^{1/2},
\end{equation}
where $L$ is the length of one side of the cube simulation box and $\mathbf{v}_t$ is the turbulent velocity field. $\int_V \, dV$ represents the volume integration over the computational domain. $c_{\rm s}$ is the sound speed at T = 10K. The numerical simulations were stopped at one million years after the formation of the first protostar.

The sink particle method, as implemented by \cite{matsumoto2015}, is used to simulate gas dynamics in a part of the molecular cloud for a long time. Gravitational collapse is thought to begin when the central density of a prestellar core reaches $n \sim 10^6 \,\rm cm^{-3}$ \citep[][]{tokuda2020}. Thus, the threshold number density $n_{\rm sink}$ for sink particle formation is set to $9 \times 10^6 \,\rm cm^{-3}$. As described in Appendix of \citet{matsumoto2015}, sink particles are created based on the following criteria: (1) the gas number density in a cell exceeds $n_{\rm sink}$, (2) gas is accreting onto the cell, (3) the gravitational potential has a local minimum, (4) gas within $r_{\rm sink}$ is gravitationally bound, and (5) no new sink particle is created within $2r_{\rm sink}$ of an existing sink particle. Once sink particles form, each sink particle accretes gas within its own radius of $r_{\rm sink} = 1.01 \times 10^3 \, \rm au$, where the gas density exceeds the threshold number density $n_{\rm sink}$ in each cell. We solve the following basic equations for hydrodynamics:
\begin{equation}
    \pdv{\rho}{t} + \nabla {\cdot} (\rho {v}) = 0,
\end{equation}
\begin{equation}
    \pdv{t}(\rho \mathbf{v}) + \nabla \cdot ( \rho \mathbf{v}\mathbf{v}^{\rm{T}} + P \mathbf{I} ) = -\rho \nabla \Phi,
\end{equation}
\begin{equation}
    \pdv{t}(\rho E) + \nabla \cdot [(\rho E + P)\mathbf{v})] = - \rho \mathbf{v} \nabla \Phi + \Gamma + \Lambda,
\end{equation}
\begin{equation}
    \nabla^2 \Phi = 4 \pi G \rho,
\end{equation}
\begin{equation}
    E = \frac{| \mathbf{v} | ^2}{2} + (\gamma - 1 )^{-1} \frac{\mathbf{P}}{\rho},
\end{equation}
where $\rho, P ,\mathbf{v}, \Phi, E, \Gamma$ and $\Lambda$ are the density, pressure, velocity, gravitational potential, total energy, the heating and cooling rates. The heating $\Gamma$ and cooling $\Lambda$ rates include the processes such as heating from chemical reactions, cooling from line emissions and energy transfer between gas and dust. We incorporate the chemical network of 11 species; $\rm H, \, H_2,\, H^-,\, H^+,\, H^+_2,\,e,\, CO,\, C\,\textsc{ii},\, O\,\textsc{i},\, O\,\textsc{ii},\, and\, O\,\textsc{iii}$. \citep[for details, see][]{fukushima2020,fukushima2021}. More than 90\% of the gas is composed of $\rm H_2$. In this study, the column density map represents $\rm H_2$ values, and proton total density is used for analyzing the physical properties of the cores.

\subsection{Implementation of Passive Tracer Particles} \label{subsec:tracer}
To focus on the trajectories of gas infalling protostars, we implement passive tracer particles into the SFUMATO code. 
As the initial condition, three million passive tracer particles are evenly distributed throughout the three-dimensional space of the simulation box, aligned with the $x$, $y$, and $z$ axes. Tracer particles are always calculated by using the information from the cells at the highest refinement level and record their own coordinates, velocities, and densities at each timestep. Based on \cite{koga2022} and \cite{moseley2023}, the advection of passive tracer particles is calculated with second-order accuracy, by the following equations:
\begin{equation}
    r^{\rm{n+1}}_i = r^{\rm{n}}_i + v_i dt + \frac{1}{2}\frac{dv_i}{dt}(dt)^2, \label{muscl}
\end{equation}
\begin{equation}
    u^{\rm{n+1}}_i = u_{\rm{near}} + \frac{x_{\rm{prtcle}} - x_{\rm{near}}}{2 \Delta x}~{\rm{minmod}} (u_{i+1}-u_i,u_i - u_{i-1}), \label{muscl2}
\end{equation}
where $r_i = (x_i, y_i, z_i)$ represents the coordinates of the $i$-th passive tracer particle, and $u_i$ is the physical quantities. $v_i$ denotes the velocities of the $i$-th passive tracer particle. The minmod function is defined as follow:
\begin{equation}
   \rm {\rm{minmod}} (a, b) = [{\rm{sgn}} (a) +{\rm{sgn}} (b)]~min(a, b).
\end{equation}
These advection calculations for the passive tracer particles are performed at every time step, recording the physical quantities. 

At the birth of a sink particle, tracer particles in cells that satisfy the accretion conditions (as described in Section \ref{subsec:settings}) are accreted onto the sink particle. Once accreted, their advection calculations are stopped. After a sink particle forms, the gas and tracer particles in cells satisfying these conditions continue to fall onto the sink particle, following the same accretion rules. These rules remain unchanged throughout the simulation.

\subsection{Terminology}\label{subsec:defcore}
\begin{figure}[htbp]
     \centering
	\includegraphics[width=\columnwidth]{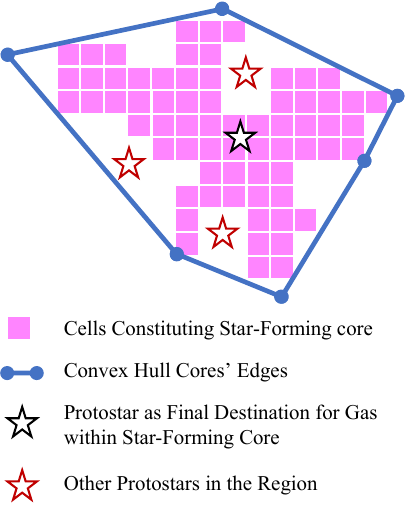} 
    \caption{Conceptual image of a star-forming core and a convex hull core. A star-forming core encloses a protostar that serves as the final destination for the gas within it. Convex hull cores and star-forming cores sometimes enclose more than one protostar.} 
    \label{conceptual_image}
\end{figure}

The precursor of stars is referred to by various terms in the field of star formation, including molecular cloud core, prestellar core, dense core, starless core, progenitor core, and star-forming core, each with different definitions. In numerical simulations at the isolated core scale, a spherically symmetric core such as Bonnor-Ebert (BE) sphere is usually assumed as the core. In numerical simulations at the clump scale, cores are identified using methods similar to observational approaches. In the following, we define the two types of cores derived from our new analysis methods, distinct from observational approaches. We analyzed the two types of cores to understand the characterization of the infalling matter onto protostars.
\begin{description}
  \item[Star-Forming Core]\mbox{}\\  
  Defined as the mass reservoir regions for protostars. The star-forming core, identified using the algorithm described in Section \ref{subsec:getcore}, represents the infalling gas distribution onto the sink particle region, highlighted in pink in Fig. \ref{conceptual_image}. The definition of a star-forming core in this study differs from that identified in observations, which relies on parameters such as gas density and velocity.
  \item[Convex Hull Core]\mbox{}\\  
  Defined as the smallest convex polyhedron that encloses the star-forming core and the surrounding gas not accreting onto a protostar, as shown by the blue lines in Fig. \ref{conceptual_image}. Analyzing convex hull cores in relation to the star-forming cores enables us to assess more detailed physical characteristics of the star-forming cores, such as porosity. 
\end{description}
We discuss findings about the star-forming core in Section \ref{sec:mass_reservoir} and about the convex hull core in Section \ref{sec:convex}.

\subsection{Core Identification Algorithm} \label{subsec:getcore}
We have constructed an algorithm to identify star-forming cores based on both passive tracer particles falling onto protostars and the density field. We identify a star-forming core as following steps.
\begin{enumerate}
    \item 
    Identify the tracer particles that will accrete onto a protostar within 0.3 million years after protostar formation. Then, trace the positions of these identified tracer particles back to the moment immediately following the formation of the protostar. Passive tracer particles record their positions and velocities at each timestep until accreting onto a protostar, enabling reconstruction of the accretion paths. The mass of the sink particle (protostar) in the data output immediately after its formation was approximately $0.1 - 0.4 \, \msun$.
    \label{step1}
    
    \item Calculate the radius $r_{i, \rm particle}$ of a sphere representing the volume occupied by each tracer particle, based on the mass densities of passive tracer particles identified in Step \ref{step1}. This calculation is based on the mass of tracer particles and the mass density of cells, using the formula:    
    \begin{equation}
        r_{i, \rm particle} = \qty( \frac{3 M_{\rm particle}}{4 \pi \rho_{\rm cell}} )^{\frac{1}{3}}, \label{rtracer}
    \end{equation}
    where ${M}_{\rm particle} = 0.001 \, \msun$ per tracer particle. This value is determined by dividing the total mass in the simulation box ($3000 \, \msun$) by the total number of tracer particles ($ 3 \times 10^6$). The tracer particles are evenly distributed throughout the simulation box at the initial state, and we assume they all have the same mass. When the mass density of cells is extremely low, the estimated volume of the cells may be inaccurate. Therefore, cells with a proton number density below $100 \,\rm cm^{-3}$ are exclude to maintain accuracy in volume distribution analysis. Define a star-forming core as a group of all cells located within radius of each tracer particle. 
    \label{step2}
\end{enumerate}

\section{star-forming cores as mass reservoirs} \label{sec:mass_reservoir}
\begin{figure}[htbp] 
     \centering
	\includegraphics[width=\columnwidth]{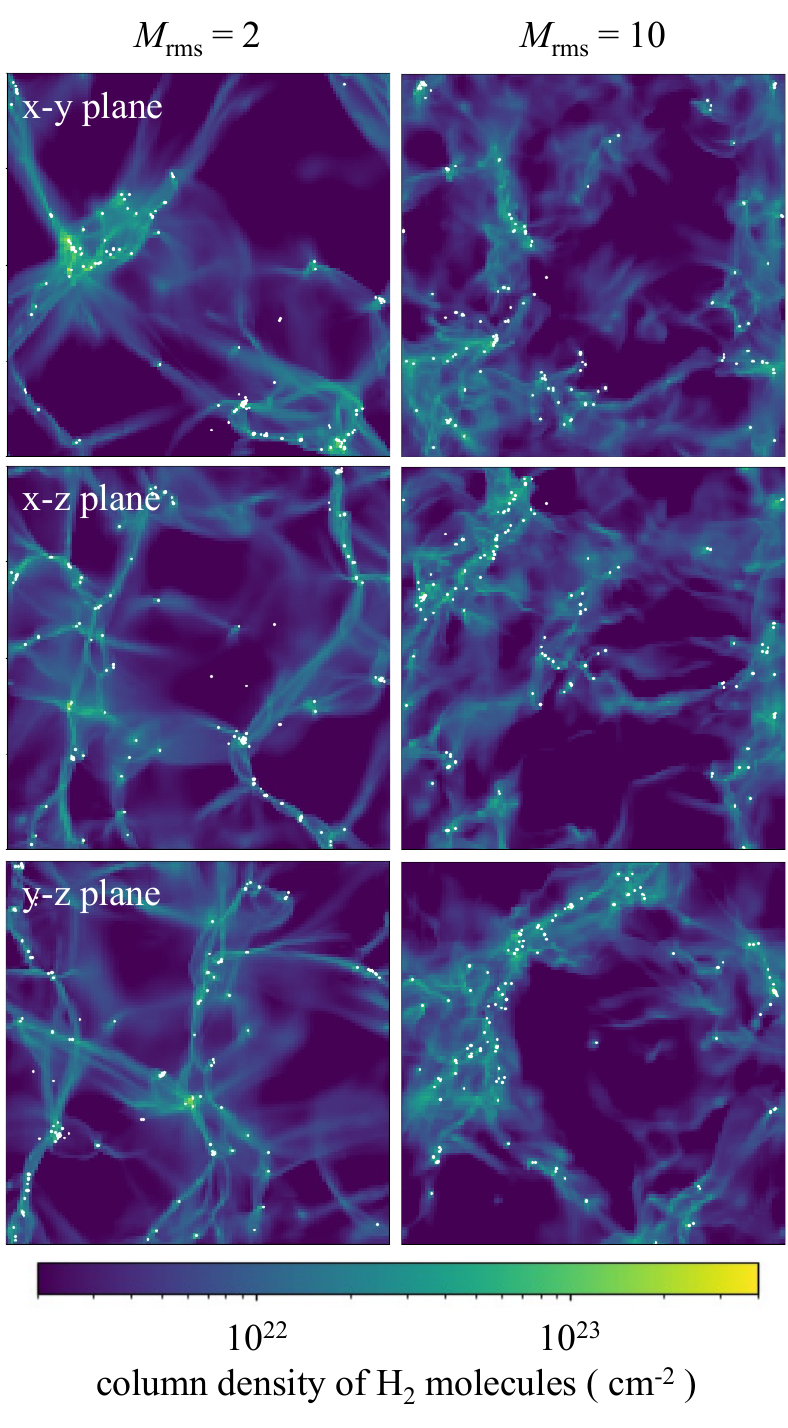}
    \caption{Column density of $\rm H_2$ molecules in each turbulence strength. White dots mark the position of the protostars (or sink particles). The color represents the column density of $\rm H_2$ molecules, with brighter colors indicating higher densities.} \label{overview}
\end{figure}

Setting initial turbulence strength as a parameter, we conducted clump-scale numerical simulations representing a region of a molecular cloud. Fig. \ref{overview} illustrates the column density of $\rm H_2$ molecules, $N_{\rm H_2}$ one million years after the formation of the first protostar.
The column density of high-density filamentary structures is approximately $10^{23}\, \rm cm^{-2}$, which is comparable to the column densities observed in low-mass star-forming regions and Bok globules, as reported in many studies \citep[e.g., ][]{alves2001, motte2001, kandori2005, palmeirim2013}. Protostars form along high-density filaments, consistent with observations showing that many molecular cloud cores are distributed along such structures \citep[e.g., ][]{andre2010, konyves2015, ladjelate2020}.

\subsection{Comparison of Stellar Mass Reservoirs and Dense Star-Forming Regions}\label{subsec:mr_morphology}
\begin{figure*}[htbp]
     \centering
	\includegraphics[width=\textwidth]{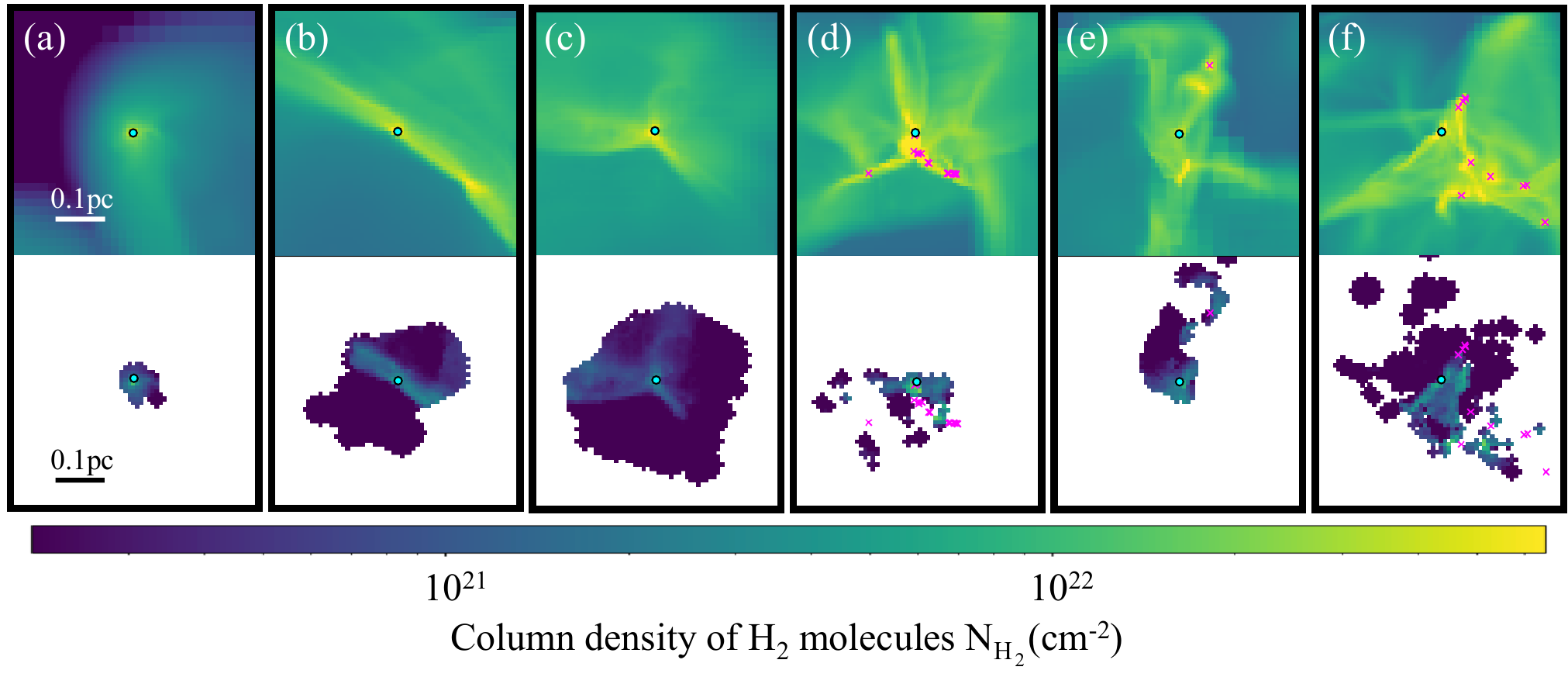} 
    \caption{Column density around the protostar (cyan dot) (top) and corresponding column density of the identified star-forming cores (bottom). The color represents the column density of $\rm H_2$ molecules, with brighter colors indicating higher densities. The magenta crosses mark the position of surrounding protostars. When the other protostars are present in the regions, as seen in (d), (e) and (f), star-forming cores form numerous small, dense clumps.}
    \label{identified_core}
\end{figure*}
As described in section \ref{sec:methods}, the star-forming cores are identified by using passive tracer particles in the numerical simulation. The identified star-forming cores are represented by the positions of the gas at the epoch of protostar formation, which accrete onto the protostars within 0.3 million years. This indicates that, in our definition, star-forming cores are completely agreement with the actual mass reservoir for the protostars. In the model with $\mathcal{M}_{\rm rms} = 2$, 264 protostars form, of which 150 star-forming cores are identified. In the model with $\mathcal{M}_{\rm rms} = 10$, 217 protostars form, of which 110 star-forming cores are identified. The total number of protostars is shown in Appendix \ref{asec:imf}. The identified star-forming cores represent only those protostars whose evolution can be traced for 0.3 million years. If a protostar forms within 0.3 million years before the end of the simulation, the corresponding star-forming core is excluded from our analysis.

\begin{figure}[htbp]
     \centering
	\includegraphics[width=\columnwidth]{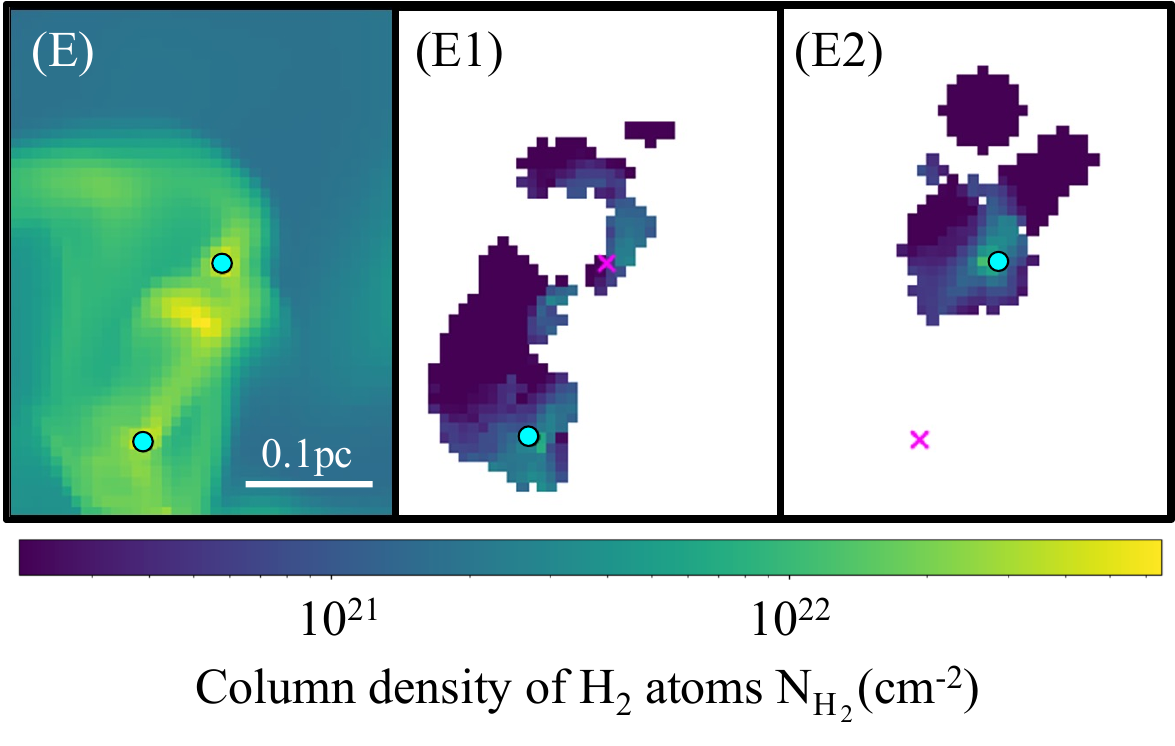} 
    \caption{Column density around the protostar (cyan dot) and corresponding column density of the identified star-forming cores, representing the gas accreting onto each protostar (cyan dot), displayed in panels (E1) and (E2). The color represents the column density of $\rm H_2$ molecules. Magenta crosses mark the positions of surrounding protostars.}
    \label{identified_core_zoom}
\end{figure}

A subset of the identified star-forming cores are shown in Fig. \ref{identified_core}. The bottom panels of Fig. \ref{identified_core} illustrate the column density of the star-forming cores, with cyan dots indicating the positions of the protostar. The top panels show the column density of the surrounding regions. Magenta crosses mark the locations of other protostars present in the same region, distinguishing them from those indicated by cyan dots. The size and shape of the mass reservoir regions of every protostar varied considerably, as seen in the morphologies of the star-forming cores in Fig. \ref{identified_core} (a) to (f). This variation seems to depend on the number of surrounding protostars and the surrounding density distribution. 

Star-forming cores that have no other protostars within 0.1 pc exhibit both radially isotropic and anisotropic density structures. As shown in Fig. \ref{identified_core} (a), isolated star-forming cores have radially isotropic density profiles. Isolated star-forming cores vary in size, ranging from radii less than 0.1 pc to more than 0.2 pc. On the other hand, as shown in Fig. \ref{identified_core} (b), there are also isolated star-forming cores that contain anisotropic density structures such as single filaments. Fig. \ref{identified_core} (c) shows an isolated star-forming core that contains a hub-filament system. Additionally, the star-forming cores also include relatively diffuse gas that is perpendicular to the dense filamentary structures in the column density maps of the surrounding regions, as shown in Fig. \ref{identified_core} (b) and (c).

As shown in Fig. \ref{identified_core} (d), (e) and (f), other protostars were also identified in the region close to the star-forming core associated with a protostar (cyan dot). These star-forming cores, defined as mass reservoir regions, are clumped and consist of numerous small, dense clusters. Despite all star-forming cores being defined as gas that will fall onto the protostars within the same 0.3 million years, their sizes differs significantly. As shown in the bottom panels of Fig. \ref{identified_core} (d), (e) and (f), the extent of the star-forming cores with closely located protostars far exceeds 0.1 pc.

In the top panel of Fig. \ref{identified_core} (e), two protostars (cyan dots and magenta crosses) are present in the column density maps of the surrounding regions including the star-forming cores. In the bottom panel of Fig. \ref{identified_core} (e), the column density map of star-forming core, which corresponds to the protostar indicated by the cyan dot, is shown. The column densities of the star-forming cores corresponding to each protostar are shown in Fig. \ref{identified_core_zoom}. Among the two star-forming cores, core E1 has a void region on the upper side, which corresponds to core E2. 

In region where protostars are clusterd, gas competitively accretes onto each protostars. Thus, when multiple other protostars are located near the star-forming core, the shape of the core as a mass reservoir can be highly complex.

\subsection{Shape in Star-Forming Cores\label{subsec:gasmass}}
To quantitatively evaluate the spatial distribution of star-forming cores at different density thresholds, 
we derived the ratio of the major to minor axes in the principal axis coordinates (see Appendix \ref{asec:shape}). The principal axis coordinates are defined based on the direction of maximum variance in the mass distribution of the star-forming core. The axis ratio of the star-forming core is defined as 
\begin{equation}
     \gamma_{\rm{core}}  = \frac{\lambda_{\text{max}}}{\lambda_{\text{min}}},
     \label{aratio}
\end{equation}
where $\lambda_{\rm max}$ and $\lambda_{\rm min}$ are the largest and smallest eigenvalues of the intertia tensor.

\begin{figure}[htbp]
     \centering
	\includegraphics[width=0.96\columnwidth]{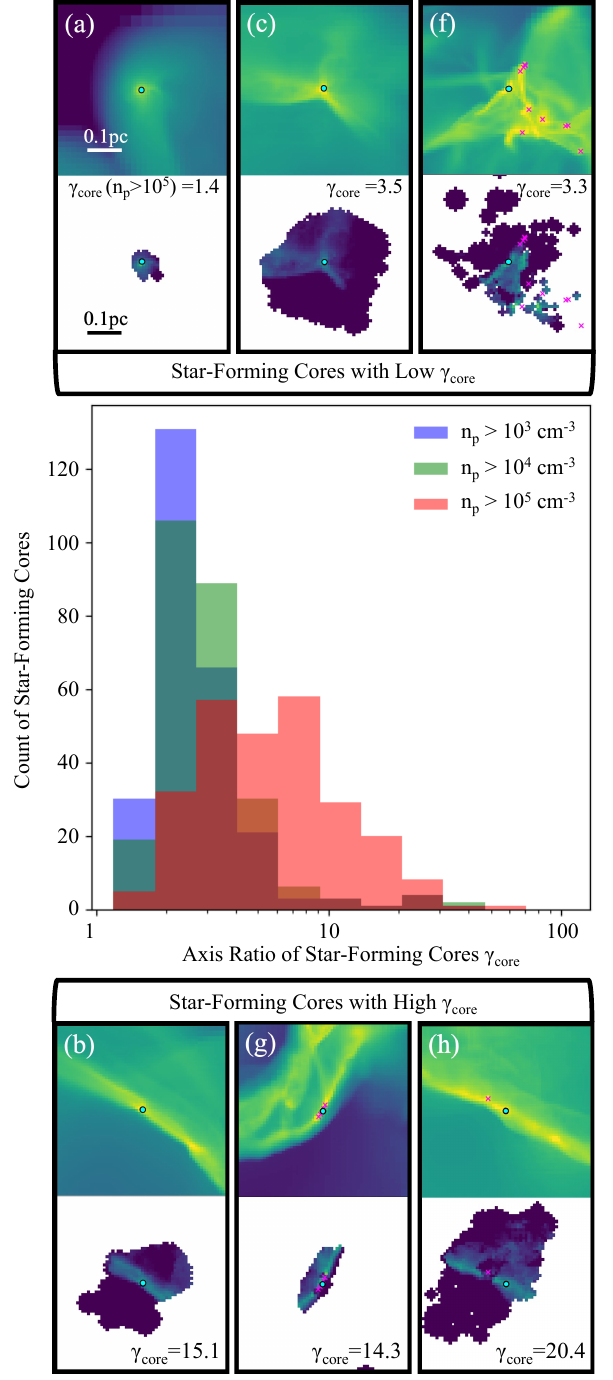} 
    \caption{Histogram of the axis ratio $\gamma_{\rm core}$ in the principal coordinates for 260 star-forming cores. The color indicates the different density thresholds for the region where the axis ratios were calculated. The larger value of $\gamma_{\rm core} (\equiv \lambda_{\rm max} / \lambda_{\rm min})$, the more asymmetric the structure becomes. In the upper part of the figure, the column density of a subset of identified star-forming cores with low $\gamma_{\rm core}$ and the column density around the protostar are shown. In the lower part of the figure, the column density of a subset of identified star-forming cores with high $\gamma_{\rm core}$ and the column density around the protostar are displayed.}
    \label{axis_ratio}
\end{figure}

Regardless of the different turbulence strengths, the axis ratio $\gamma_{\rm{core}}$ was calculated for the 260 star-forming cores. Fig. \ref{axis_ratio} shows the distribution of $\gamma_{\rm{core}}$ for each density threshold within the star-forming cores. The axis ratio $\gamma_{\rm{core}}$ for regions where proton number densities $n_{\rm p} > 10^3\, \rm cm^{-3}$ and $n_{\rm p} > 10^4\, \rm cm^{-3}$ within the star-forming cores is mainly distributed between 1 and 5, suggesting that these regions are close to spherical in shape. In contrast, the axis ratio $\gamma_{\rm{core}}$ for regions where $n_{\rm p} > 10^5\, \rm cm^{-3}$ mostly falls within the range of 1 to 30. In regions with higher density, the distribution of axis ratio values becomes broader. This indicates that the shape of star-forming core regions with high density varies from spherical to elongated or flattened structures, compared to those with low density. While a hierarchical pattern might hold in lower-density regions, the increased variety in high-density regions differs from the expected hierarchical structure.

In regions where $n_{\rm p} > 10^5 \, \rm{cm}^{-3}$, star-forming cores with an axis ratio $\gamma_{\rm core}$ exceeding 10 exhibit significant anisotropy. As shown in the bottom panels of Fig. \ref{axis_ratio}, the star-forming cores with $\gamma_{\rm core} > 10$, corresponding to (b), (g) and (h), include a single filamentary structure, regardless of the presence of surrounding protostars. Star-forming cores with relatively small axis ratios exhibit minimal anisotropy, even in high-density regions. As shown in the top panels of Fig. \ref{axis_ratio}, these cores include radially isotropic density profiles or include multiple filamentary structures. Thus, the broader distribution of axis ratios indicates that not all cores exhibit increased anisotropy within their high-density regions.

\section{Impact of Turbulence on Star-Forming Cores: A Convex Hull Analysis \label{sec:convex}} 
\subsection{Analysis of Convex Hull Core and Filling Factor in Star-Forming Cores \label{subsec:convex}}
\begin{figure}[htbp]
     \centering
	\includegraphics[width=\columnwidth]{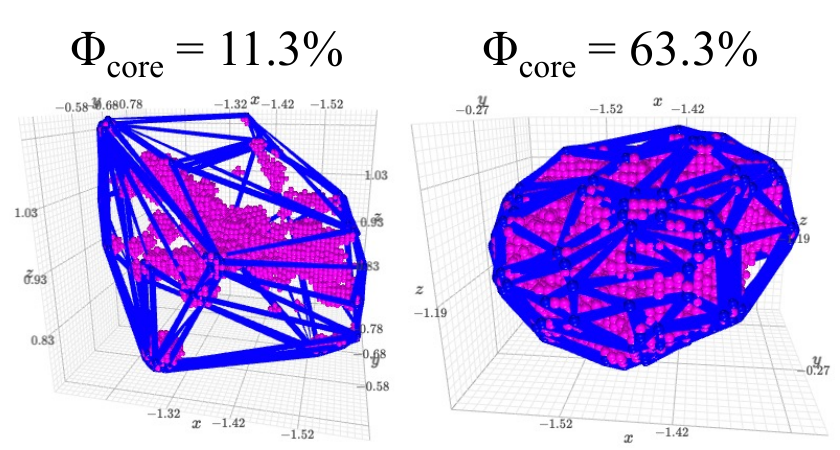} 
    \caption{Two examples of star-forming cores as mass reservoirs (pink dots) and their including convex hull cores’ edges (blue lines). The left panel shows a core with a low filling factor (11.3\%), while the right panel shows a core with a high filling factor (63.3\%).}
    \label{convex_hulls}
\end{figure}

To treat star-forming cores as continuous structures, the smallest convex polyhedron that encloses the identified star-forming core regions, known as a convex hull, was determined. As shown in Fig. \ref{identified_core}, star-forming cores, defined as mass reservoirs, are clumped and consist of numerous small, dense clusters when other protostars are closely located. Using the quick hull method, convex hulls were calculated for 260 star-forming cores and these regions were defined as convex hull cores, as illustrated in Fig \ref{conceptual_image}. Fig. \ref{convex_hulls} illustrates two examples of the three-dimentsional extent of the star-forming cores and their including convex hull cores. 
The pink dots represent the cells constituting the star-forming cores, and the blue edges indicate the convex hull cores that enclose the star-forming cores. The convex hull cores have a larger volume than the identified star-forming cores and include gas that does not fall onto the protostars.

The filling factor of a core is defined using the total volume of the cells included in the star-forming core identified in Section \ref{subsec:getcore} ($V_{\rm core}$) and the total volume of the cells included in the convex hull core ($V_{\rm hull}$) as
\begin{equation}
     \phi_{\rm core} (\%) = \frac{V_{\rm core}}{V_{\rm hull}} \times 100.
     \label{def_fill}
\end{equation}
The filling factor $\phi_{\rm core}$ indicates the the distribution of gas within the star-forming core and represents how centrally concentrated the core is. A filling factor close to 100\% means that the star-forming core is densely packed within the convex hull core, as illustrated in the right panel of Fig. \ref{convex_hulls}. Conversely, a low filling factor indicates that the convex hull core contains small, clumpy structures with many voids, as shown in the left panel of Fig. \ref{convex_hulls}. The filling factors $\phi_{\rm core}$ for 260 star-forming cores were calculated, and the correlation between these factors and the physical properties of the star-forming cores are presented in the following sections.

\subsection{Correlations Between Filling Factor and Core Properties}\label{subsec:properties}
\begin{figure}[htbp]
    \centering
	\includegraphics[width=\columnwidth]{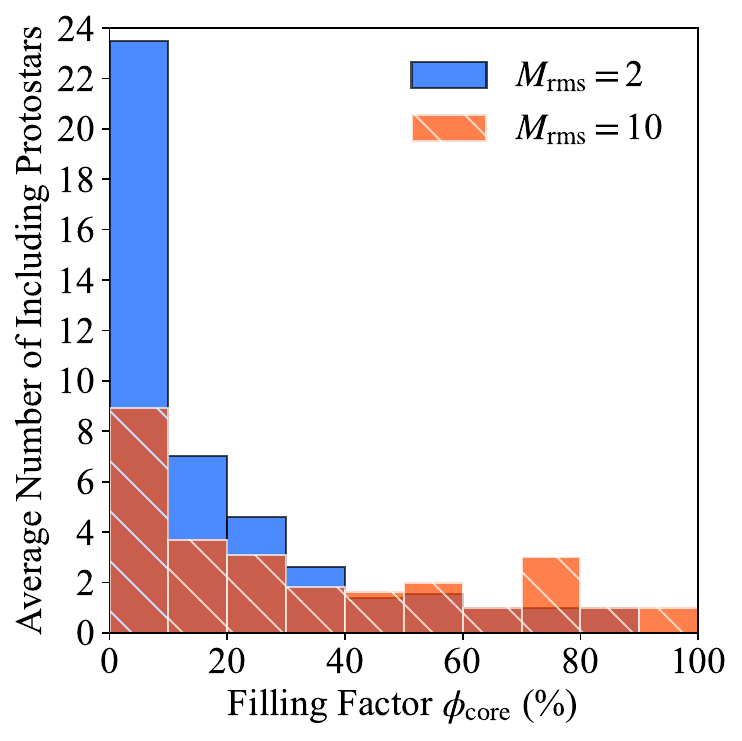} 
    \caption{Average number of protostars within the convex hull volume in each filling factor bin against the filling factor of star-forming cores immediately following protostar formation for models with $\mathcal{M}_{\rm rms} = 2$ (blue bar) and 10 (orange bar).}
    \label{fill-nps}
\end{figure}

The average number of protostars contained within the convex hull core, corresponding to each filling factor value, is shown in Fig. \ref{fill-nps}.
For example, convex hull cores without nearby protostars have only one protostar as the final accretion point for gas within the star-forming core. In the model with $\mathcal{M}_{\rm rms} = 2$ (blue), convex hull cores with $\phi_{\rm core} \ge 40\%$ have an average of around one protostar. This indicates that protostars in convex hull cores with high filling factors are isolated, without any nearby protostars. In contrast, convex hull cores with $\phi_{\rm core} \le 40\%$ have at least two protostars. This suggests that the protostar in the convex hull cores with low filling factor values is not isolated, with at least one additional protostar, as shown in Fig \ref{conceptual_image}. As the filling factor $\phi_{\rm core}$ decreases, the average number of protostars tends to increase. 

In the model with $\mathcal{M}_{\rm rms} = 10$ (orange), convex hull cores show a similar trend; surrounding protostars increases as the filling factor $\phi_{\rm core}$ decreases, as seen in the model with $\mathcal{M}_{\rm rms} = 2$. For filling factor ranges $40\% \le \phi_{\rm core} \le 100\%$, the average number of protostars is between one and three, indicating that most convex hull cores do not have any protostars other than the one as the terminal point for gas accretion within the star-forming core. In the model with $\mathcal{M}_{\rm rms} = 10$, convex hull cores with extremely low filling factors, such as $0\% \le \phi_{\rm core} \le 40\%$, contain fewer surrounding protostars compared to those in the model with $\mathcal{M}_{\rm rms} = 2$.

\begin{figure*}[htbp]
    \centering
    \includegraphics[width=\textwidth]{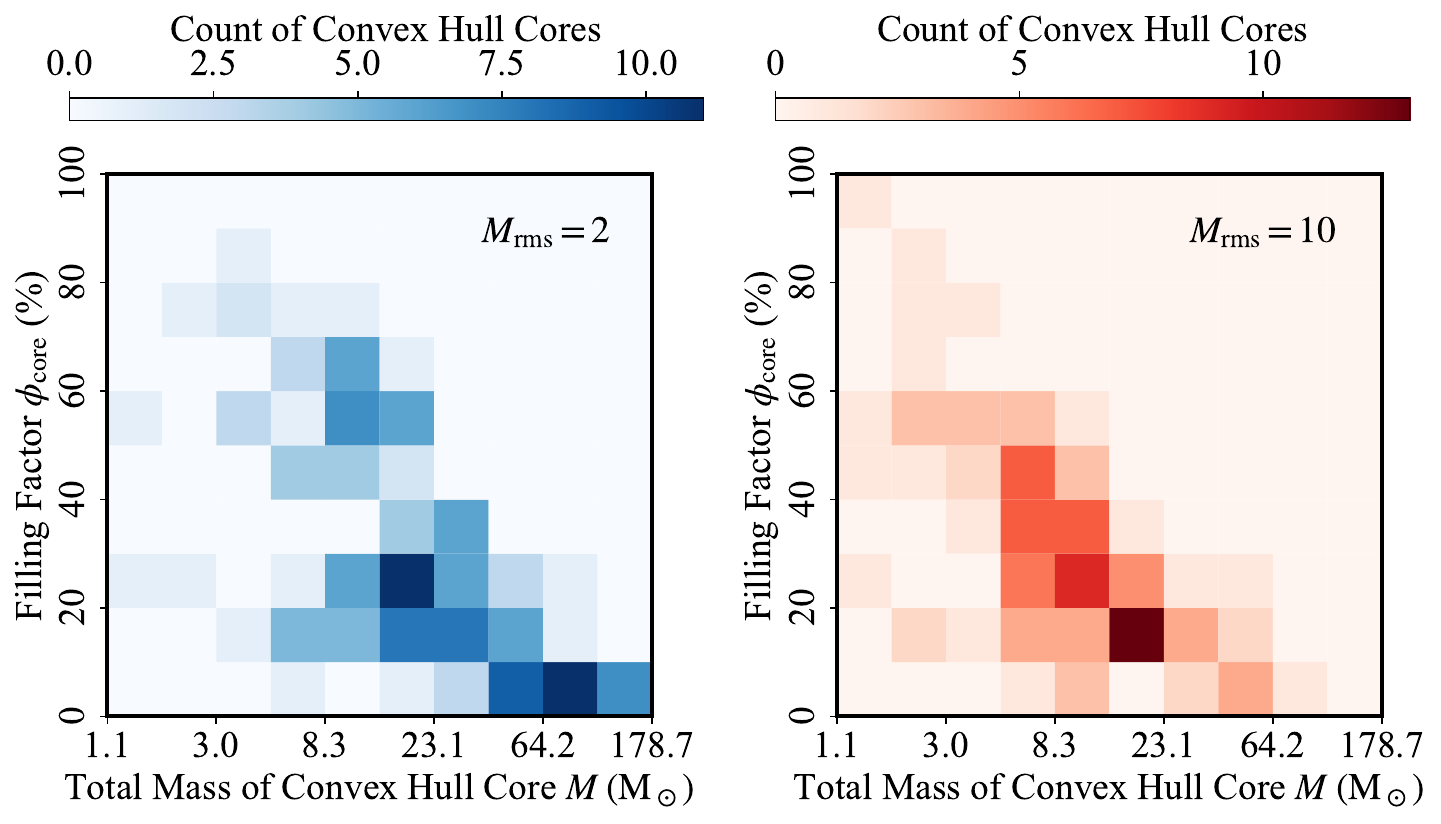} 
    \caption{Correlation between the mass within the convex hull volume ($\rm log_{10}(M/M_\odot)$) and the filling factor (\%). The left panel corresponds to the model with $\mathcal{M}_{\rm rms} = 2$, while the right panel corresponds to the model with $\mathcal{M}_{\rm rms} = 10$. The color scale indicates the number of convex hull cores in each bin.}
    \label{fill-mass}
\end{figure*}

Fig. \ref{fill-mass} presents the correlation between the filling factor of each convex hull cores and its total mass. 
The mass of each convex hull core is the sum of the gas mass and the protostar masses contained within it. The convex hull core mass exceeds the mass of the identified star-forming cores. As shown in the left panel of Fig. \ref{fill-mass}, in the model with $\mathcal{M}_{\rm rms} = 2$, convex hull cores with the filling factor $\phi_{\rm core} \le 50\%$ have the mass range $1.1 \le M \, (\msun) \le 178.7$. Convex hull cores with the filling factor $\phi_{\rm core} \ge 50\%$ have masses up to $23.3 \, \msun$. Overall, the larger the masses of the convex hull cores, the lower their filling factors. Some low-mass convex hull cores have small filling factors; however, no massive convex hull cores have large filling factors. All massive convex hull cores exhibit low filling factors, suggesting that massive cores are associated with cluster-like star formation, rather than the formation of a single massive star.

As shown in the right panel of Fig. \ref{fill-mass}, the model with $\mathcal{M}_{\rm rms} = 10$ exhibits a similar correlation between the filling factor of the convex hull cores and their mass as in the model with $\mathcal{M}_{\rm rms} = 2$. Convex hull cores with $M \ge 8.3 \, \msun$ exhibit the filling factors lower than $60 \%$. Also in the model with $\mathcal{M}_{\rm rms} = 10$, no massive cores have large filling factors. In the $\mathcal{M}_{\rm rms} = 2$ model, 84\% of the convex hull cores have masses above $8 \, \msun$, while in the $\mathcal{M}_{\rm rms} = 10$ model, this fraction decreases to 59\%. This indicates that the convex hull cores in the $\mathcal{M}_{\rm rms} = 10$ model tend to be lower in mass.

\begin{figure}[thbp]
     \centering
	 \includegraphics[width=\columnwidth]{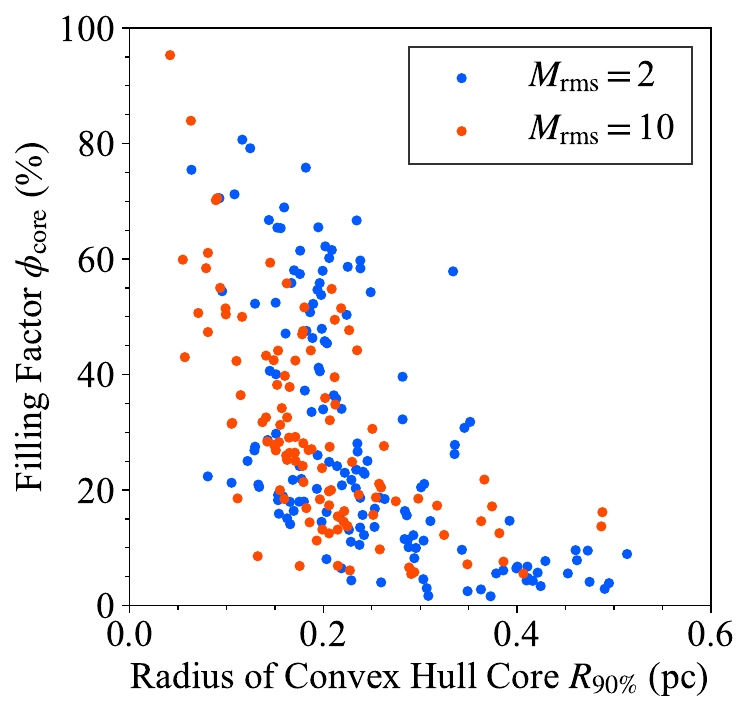} 
     \caption{Correlation between the radius enclosing 90\% of the total mass of the convex hull core and the filling factor of convex hull cores. The blue dots correspond the model with $\mathcal{M}_{\rm rms} = 2$, and the red dots represent the model with $\mathcal{M}_{\rm rms} = 10$.}
    \label{fill-extent}
\end{figure}

In addition to the correlation between mass and filling factors, Fig. \ref{fill-extent} shows the correlation between the three-dimensional extent and filling factors of the convex hull cores.
The distance from the protostar, as the final accretion point for star-forming gas, to each cell within the convex hull core was calculated, and the radius enclosing 90\% of the convex hull core mass was determined. This radius is defined as $R_{90\%}$. As shown in Fig. \ref{fill-extent}, the convex hull cores in the $\mathcal{M}_{\rm rms} = 2$ model have a radius range of $0.06 \le R_{90\%} \, ({\rm pc}) \le 0.51$, and those in the $\mathcal{M}_{\rm rms} = 10$ model exhibit a similar range of $0.04 \le R_{90\%} \, ({\rm pc}) \le 0.49$. The similarity in core radii between both models suggests that differences in turbulence strength do not significantly affect the radii of convex hull cores. The average radius of the convex hull cores in both the $\mathcal{M}_{\rm rms} = 2$ and $\mathcal{M}_{\rm rms} = 10$ models is 0.22 pc.

To investigate the impact of turbulence on star-forming cores, we analyzed the stability of the convex hull cores. The energy ratio $\alpha_{\rm virial}$ is derived from the physical quantities of the convex hull core, and is defined as
\begin{equation}
    \alpha = \frac{E_{\rm k} + E_{\rm t}}{| E_{\rm g} |}. \label{eq:virial}
\end{equation}
In equation \eqref{eq:virial}, $\alpha \le 1$ means that the gravitational energy is dominant, implying a gravitationally unstable star-forming core. However, $\alpha \ge 1$ indicates that the kinetic energy or the thermal energy is dominant. This energy condition to evaluate the stability of the convex hull cores is used in \cite{smith2009}, \cite{kong2015} and \cite{pelkonen2021}. Equations for deriving each energy are given by 
\begin{align}
    E_{\rm g} &= - \sum_{i=1}^n \sum_{j \neq i}^n \frac{G m_{i} m_{j}}{r_{ij}} \label{eq:virial:g}, \\ 
    E_{\rm k} &= \sum_{i=1}^n \frac{1}{2} m_{i} (v_i - \bar{v})^2 \label{eq:virial:k}, \\ 
    E_{\rm t} &= \frac{3}{2} V_{\rm total} k_{\rm B} \sum_{i=1}^n n_{{\rm p} , i} T_i. \label{eq:virial:t}
\end{align}
Equation \eqref{eq:virial:g} presents the gravitational energy $E_{\rm g}$, which is calculated by performing a double integral over the mass of each cell within a convex hull core and the masses of the protostars, denoted by $m$, and the relative distance $r$ between protostars and each cell. Equation \eqref{eq:virial:k} shows the kinetic energy $E_{\rm k}$, which is calculated by taking the mass-weighted velocity dispersion of a convex hull core as given by
\begin{equation}
    \bar{v} = \sum_{i=1}^n \qty(\frac{m_i}{\sum_{j=1}^n m_j} v_i).
\end{equation}
The thermal energy $E_{\rm t}$ is derived from equation \eqref{eq:virial:t}. $V_{\rm total}$ represents the volume of a convex hull core. $n_{\rm p}$ and $T$ are the proton number density and the temperature of each cell within a convex hull core. The equations to evaluate the kinetic and thermal energies are the same as those adopted in \cite{pelkonen2021}.

\begin{figure}[htbp]
     \centering
	\includegraphics[width=\columnwidth]{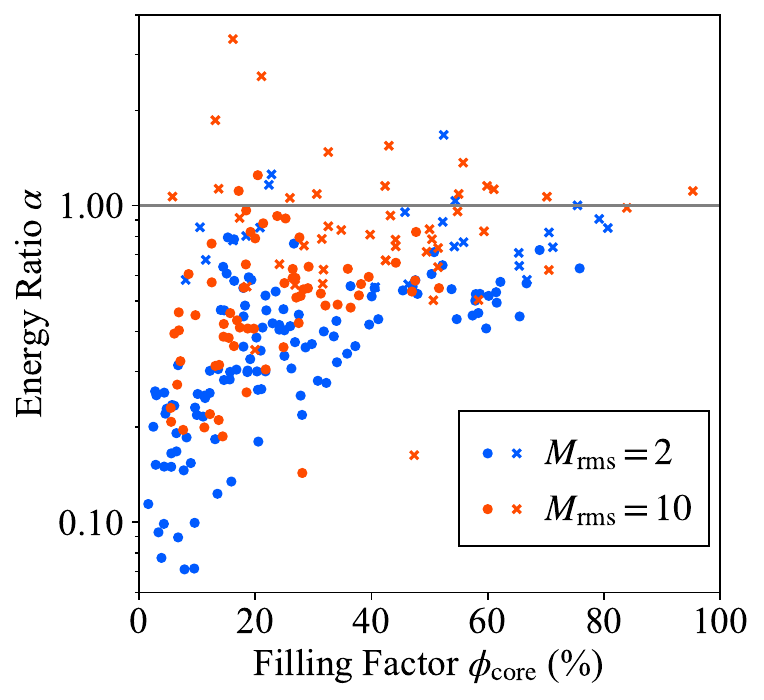} 
     \caption{Energy ratios for different filling factors of convex hull cores. Red dots and crosses represent $\mathcal{M}_{\rm rms} = 2$, and blue dots and crosses represemt $\mathcal{M}_{\rm rms} = 10$. Dots indicate convex hull cores with masses of $ \ge 8 \, \msun$, and crosses indicate those with masses $< 8 \, \msun$. In this figure, five convex hull cores with energy ratios below 0.05 are excluded.}
    \label{virial}
\end{figure}
In Fig. \ref{virial}, 97\% of the convex hull cores in the model with $\mathcal{M}_{\rm rms} = 2$ and 84\% of those in the model with $\mathcal{M}_{\rm rms} = 10$ have energy ratios $\alpha \le 1$. Thus, most of convex hull cores in both the $\mathcal{M}_{\rm rms} = 2$ and $\mathcal{M}_{\rm rms} = 10$ models are gravitationally bound when considering all protostars contained within the convex hull cores. However, some convex hull cores are gravitationally unbound, particularly in the $\mathcal{M}_{\rm rms} = 10$ model, where 16 \% of the convex hull cores are unbound. A large fraction of convex hull cores with energy ratios greater than or close to unity are relatively low-mass, with masses $< 8 \, \msun$.

\section{Discussion} \label{sec:discussion}
\subsection{Cloud to Core Structure: Qualitative Comparison with Observations }\label{subsec:structure_previous}
In this study, as shown in Fig. \ref{identified_core}, the star-forming cores identified as mass reservoirs for protostars are present in various shapes and are located within isolated clumps, as well as along filamentary and hub-filamentary structures. These features are consistent with some observational findings in star-forming regions \citep[e.g., ][]{andre2014,konyves2010,konyves2015, kumar2020, tokuda2023}. However, our identified star-forming core regions include not only the high-density gas structures likely identified in observations but also diffuse gas located outside the filamentary structures. Therefore, our simulations suggest that the protostars may accrete gas from a wider region beyond the high-density regions identified in observations. Note that the sizes of cores identified in observations strongly depend on observational settings and analysis techniques, which makes consistent definitions across observations challenging. To make quantitative comparisons with observed cores and evaluate whether protostars are accreting gas from wider regions, analyses such as synthetic observations are required.

Some of the star-forming cores identified by our method, as shown in Fig. \ref{identified_core} (b), include both high-density filamentary structures and lower density gas extending perpendicular to these filaments. This structure implies that mass is being accreted onto the dense cores and protostars through the mass accretion onto the filaments from the low-density regions. This finding aligns with recent observations \citep{palmeirim2013, shimajiri2019}, which suggest that material may be accreting onto filaments through lower density striations perpendicular to the filaments. 

For simplicity, we have ignored magnetic fields in this study. Since the orientation of magnetic fields can influence filament formation \citep[e.g., ][]{seifried2015}, it is challenging to directly and quantitatively compare the results of this study with observations on the mass supply to dense cores through mass accretion onto filaments. In the future, it will be necessary to investigate the impact of mass accretion onto filaments on star-forming cores more quantitatively through numerical simulations with magnetic fields.

In regions where protostars are closely clustered, the structure of the cores identified in this study shows significant differences from those identified in observations. As shown in Fig. \ref{identified_core} (d), (e), (f), in cluster-forming regions, the star-forming cores identified in this study consist of many small gas clumps. This clumpy structure likely results from multiple stars selectively accreting gas within the same region (see Fig. \ref{identified_core_zoom}). Similarly, \cite{david2023} reported that a large fraction of cores overlap with multiple others, which supports our findings that mass reservoirs for protostars exhibit a clumpy structure. In contrast, observational studies identify dense cores as singular entities due to limitations in identification methods, which focus on high-density regions rather than the full extent of the mass reservoirs \citep[e.g., ][]{takemura2021a,takemura2021b}. Therefore, in clustered regions, the cores identified in observations likely differ greatly from the actual mass reservoirs, as they fail to capture the full, fragmented structure and only identify the densest regions.

\subsection{Filling Factor and Cluster-Forming Region}\label{subsec:cluster_previous}
In our study, we defined the filling factor of convex hull cores to investigate the physical characteristics of cluster-forming regions. We found that convex hull cores with lower filling factors tend to contain more protostars, indicating they are associated with cluster-forming regions. \cite{david2023} also calculated core filling fractions from simulation data without sink particles, finding a distribution of filling factors peaking at 25\%, which aligns with our results for the model with $\mathcal{M}_{\rm rms} = 10$.

Our results also revealed that the number of convex hull cores increases with lower filling factors, regardless of turbulence strength. This trend indicates that convex hull cores with lower filling factors correspond to cluster-forming regions. Convex hull cores with lower filling factors tend to have larger mass and size compared to those with higher filling factors (see Fig. \ref{fill-mass} and Fig. \ref{fill-extent}). This trend showed little dependence on turbulence strength. Thus, the filling factor may serve as an important indicator of the physical properties in cluster-forming regions.

As shown in Fig. \ref{fill-mass}, although low-mass convex hull cores with low filling factors exist, no high-mass convex hull cores with high filling factors are found. Massive convex hull cores generally have low filling factors and host multiple stars (see also Fig. \ref{fill-nps}). This result supports the star formation scenario where massive cores give birth to multiple stars that selectively accrete gas \citep[e.g., ][]{bonnell2006,wang2010}, rather than the scenario where a single massive core forms only a single star \citep[e.g., ][]{mckee2003}.

\subsection{Comparison of Gravitational stability in Cores with other Studies}\label{subsec:energy_previous}
Fig. \ref{virial} shows that most of convex hull cores are gravitationally bound. This finding aligns with the general understanding that pre-stellar cores are gravitationally bound as they form protostars through gravitational core collapse. Similar trends have been seen in other numerical simulation studies, such as \cite{smullen2020} and \cite{offner2022}. Using dendrograms, \cite{smullen2020} reported a decrease in virial numbers in leaves identified from numerical simulation data as they evolve. \cite{offner2022} identified cores in Mach 2 turbulence simulations, also using dendrograms, and classified them into three phases with machine learning techniques. Their findings similarly demonstrated that cores approach the collapse phase as the turbulence within cores weakens. Thus, the cores become increasingly gravitationally bound. However, unlike the cores identified in these previous studies, the convex hull cores in this study include not only the star-forming cores as mass reservoirs but also the surrounding gas and protostars. Therefore, the entire region encompassing the mass reservoir, surrounding gas, and protostars is gravitationally bound, suggesting that several stars are forming within it. This invokes the competitive accretion model \citep[e.g., ][]{bonnell2006} or the clump-fed model \citep[e.g., ][]{smith2009, wang2010}.

Fig. \ref{virial} also shows that some of convex hull cores have energy ratios greater than unity, suggesting that they are not gravitationally bound. In the $\mathcal{M}_{\rm rms} = 10$ model, gravitationally unbound cores make up 16\% of the total, which is higher than in the $\mathcal{M}_{\rm rms} = 2$ model (3\%). \cite{pelkonen2021} also found that progenitor cores, identified using the clumpfind algorithm \citep{padoan2007} in their simulations, are not gravitationally bound, supporting the inertial-inflow model \citep{padoan2020}. Thus, some of the gravitationally unbound convex hull cores may be explained by the inertial-inflow model. 

A large fraction of the convex hull cores with energy ratios greater than or close to unity have masses below $8 \, \msun$. As shown in Fig. \ref{fill-mass}, the $\mathcal{M}_{\rm rms} = 10$ model has a higher proportion of low-mass convex hull cores compared to the $\mathcal{M}_{\rm rms} = 2$ model, with more gravitationally unbound cores. This suggests that differences in turbulence strength may influence the mass of the cores, which in turn could affect their gravitational binding.

Observational studies have reported that many dense cores are gravitationally bound, supporting the idea that gravitational collapse is a dominant mechanism in star formation. However, some studies, such as \cite{chen2019} and \cite{kerr2019}, reported more than ten cores that are gravitationally unbound and primarily confined by ambient pressure. \cite{singh2021} further suggested that certain systematic errors in observational data could artificially lower the observed virial parameters of cores, making them appear more bound than they actually are. These observations indicate that the energy ratio of star-forming cores and consequently the star formation scenario may vary somewhat depending on the turbulence strength.

\subsection{Caveats and Implications for Future Research}\label{subsec:caveats}
The strength of this study lies in the introduction of a new method for identifying star-forming cores as mass reservoirs for protostars, and in revealing the differences in the physical properties of cores based on turbulence strength. In this study, we ignored a magnetic field for the initial cloud and radiative feedback from massive stars to focus on the identification of mass reservoirs across different turbulent strengths. These stellar feedback mechanisms are critical as they significantly influence the evolution of star-forming cores and the determination of stellar mass. In clump-scale numerical simulations, protostellar outflows are often modeled as a constant fraction of the inflow rate. However, the mass ejection rate from the protostellar outflow varies considerably depending on factors such as magnetic field strength and the alignment between the disk and magnetic field, making simple models insufficient \citep[e.g., ][]{machida2013, hirano2020}. In the future studies, we will address the impact of magnetic fields on core formation and investigate how protostellar outflows affect the evolution of protostellar cores. Additionally, we will study how radiative feedback from massive stars influences the evolution of star-forming cores.

Our simulations considered only initial turbulence without additional turbulence driving and did not include feedback processes. For the model with $\mathcal{M}_{\rm rms} = 2$, the overall simulation time is $ \sim 2.7 \, \rm Myr$, which is 2.2 times the free-fall time. For $\mathcal{M}_{\rm rms} = 10$, the overall simulation time is $\sim 1.7 \, \rm Myr$, which is 1.5 times the free-fall time. The turbulence turnover time, defined as the simulation box scale ($L=4 \, \rm pc$) divided by the turbulent velocity scale ($v=\mathcal{M}_{\rm rms} \, C_{\rm s}$), is $\sim 10 \, \rm Myr$ for $\mathcal{M}_{\rm rms} = 2$ and $\sim 2.1 \, \rm Myr$ for $\mathcal{M}_{\rm rms} = 10$. Therefore, the initial turbulence should gradually decays over these timescales, leading the system to undergo large-scale gravitational collapse. Note that the free-fall time is derived from the initial uniform density and is $\sim 1.2 \, \rm Myr$. This setup has also been adopted in some studies \citep[e.g.,][]{gong2015, smullen2020} and may reflect star-forming regions where turbulence is driven once by a single supernova event.

Turbulence driving, defined as the injection of kinetic energy into the simulation box during the simulation to sustain turbulence, is considered important in star formation studies. \citet{mckee1977} estimated that supernova explosions occur approximately every 1 Myr on average in our galaxy. \citet{inutsuka2015} discussed that such events could drive multiple compression events, influencing molecular cloud formation. \citet{pan2016} also showed that turbulence driven by multiple supernova explosions contributes to energy injection and the formation of high-density structures. Therefore, simulations with turbulence driving have been conducted in some other studies \citep[e.g., ][]{maclow1999, federrath2010, padoan2020}. The presence or absence of turbulence driving may affect the physical properties of star-forming cores. For instance, simulations with only initial turbulence \citep[][and this study]{smullen2020} indicate that many cores are gravitationally bound, whereas simulations with turbulence driving \citep{pelkonen2021} suggest that many cores are gravitationally unbound. Thus, our results regarding whether the majority of star-forming cores are gravitationally bound may change if turbulence driving is considered during the simulation. In such a case, we expect an increase in the number of unbound cores and star formation driven by the inertial-inflow model \citep[e.g., ][]{padoan2020}. Note that to quantitatively evaluate these possibilities, we need to perform simulations incorporating turbulence driving and feedback processes.

\section{Summary} \label{sec:summary}
We conducted numerical simulations on star formation within molecular clouds, focusing on star-forming cores as the mass reservoir for each protostar in environments with varying turbulence strength. To trace gas motion, three million passive tracer particles are implemented. We constructed an algorithm to investigate star-forming cores as mass reservoirs for protostars, based on the tracer particles falling onto protostars, and identified 260 star-forming cores. By comparing the identified star-forming cores with convex hull cores, we defined the filling factor of the convex hull cores, and investigated its correlation with core properties, including the number of protostars, mass, size, and gravitational stability. Our results are summarized as follows:

\begin{enumerate}
\item Star-forming cores, which are defined as mass reservoirs for protostars, exhibit a variety of structures, including spherical, filamentary, and hub-filament forms. As the density of the star-forming cores increases, the distribution of axis ratios in the core's principal axis coordinates becomes broader. Star-forming cores with filamentary structures often include diffuse gas extending perpendicular to the filaments. The average radius of the convex hull cores is 0.22 pc. This result indicates that mass accretion onto filaments and mass inflow from the filaments to cores and protostars occur simultaneously.

\item Star-forming cores have substantially different gas density structures depending on the presence of nearby protostars. Our findings reveal that the mass reservoir regions of star-forming cores do not necessarily align with high-density areas. In regions with multiple nearby stars, gas selectively accretes onto protostars, leading to star-forming cores that are clumped and composed of numerous small, dense clusters. Thus, in cluster-forming region, mass reservoirs for protostars may not be easily identified as singular entities but instead appear as clumpy, fragmented structures.

\item A significant correlation was found between the filling factor of convex hull cores and their properties such as mass and radius. The results indicate that, regardless of turbulence strength, convex hull cores with lower filling factors tend to contain more protostars and have larger masses and sizes. This implies that the mass reservoirs in clustered regions are both more massive and more extended. The absence of high filling factor, massive convex hull cores implies that in massive cores, gas selectively accrete onto nearby protostars. The filling factor of cores could serve as an indicator of whether a star-forming region is isolated or clustered, and may also provide a clue to constraining star formation scenarios.

\item We calculated the energy ratios of convex hull cores and found that 97\% of the cores in the $\mathcal{M}_{\rm rms} = 10$ model and 84\% in the $\mathcal{M}_{\rm rms} = 2$ model are gravitationally bound. Notably, convex hull cores with high filling factors are also gravitationally bound, which aligns with the competitive accretion or clump-fed models. However, some cores are gravitationally unbound, particlularly the lower-mass ones. In the $\mathcal{M}_{\rm rms} = 10$ model, there are more low-mass, unbound cores compared to the $\mathcal{M}_{\rm rms} = 2$ model. The unbound cores identified in this study may be explained by the inertial-inflow model, which suggests that protostars grow through mass inflows from gravitationally unbound core regions. These findings indicate that differences in turbulence strength influence the mass of the cores as mass reservoirs, which in turn affects their gravitational stability.
\end{enumerate}

\section*{acknowledgements} \label{sec:acknowledgements}
We thank the referee for their very useful comments and suggestions, which significantly improved this paper. We are also grateful to Yoshito Shimajiri for his helpful comments. Furthermore, we deeply appreciate Tomoaki Matsumoto for his great contribution to the code development. Numerical simulations in this paper were conducted by the Cray XC50 (Aterui II) at the Center for Computational Astrophysics of National Astronomical Observatory of Japan. This work was supported in part by MEXT/JSPS KAKENHI Grant Number JP23K13139 (HF), JP21H00046, JP21K03617 (MNM), JP20H05645, JP21H00049, JP21K13962 (KT), and JST SPRING, Grant Number JPMJSP2136(SN). This work was also supported by a NAOJ ALMA Scientific Research grant (No. 2022-22B).

\vspace{5mm}

\appendix
\restartappendixnumbering

\section{mass distribution of protostars} \label{asec:imf}
\begin{figure}[htbp]
    \centering
    \includegraphics[width=0.7\columnwidth]{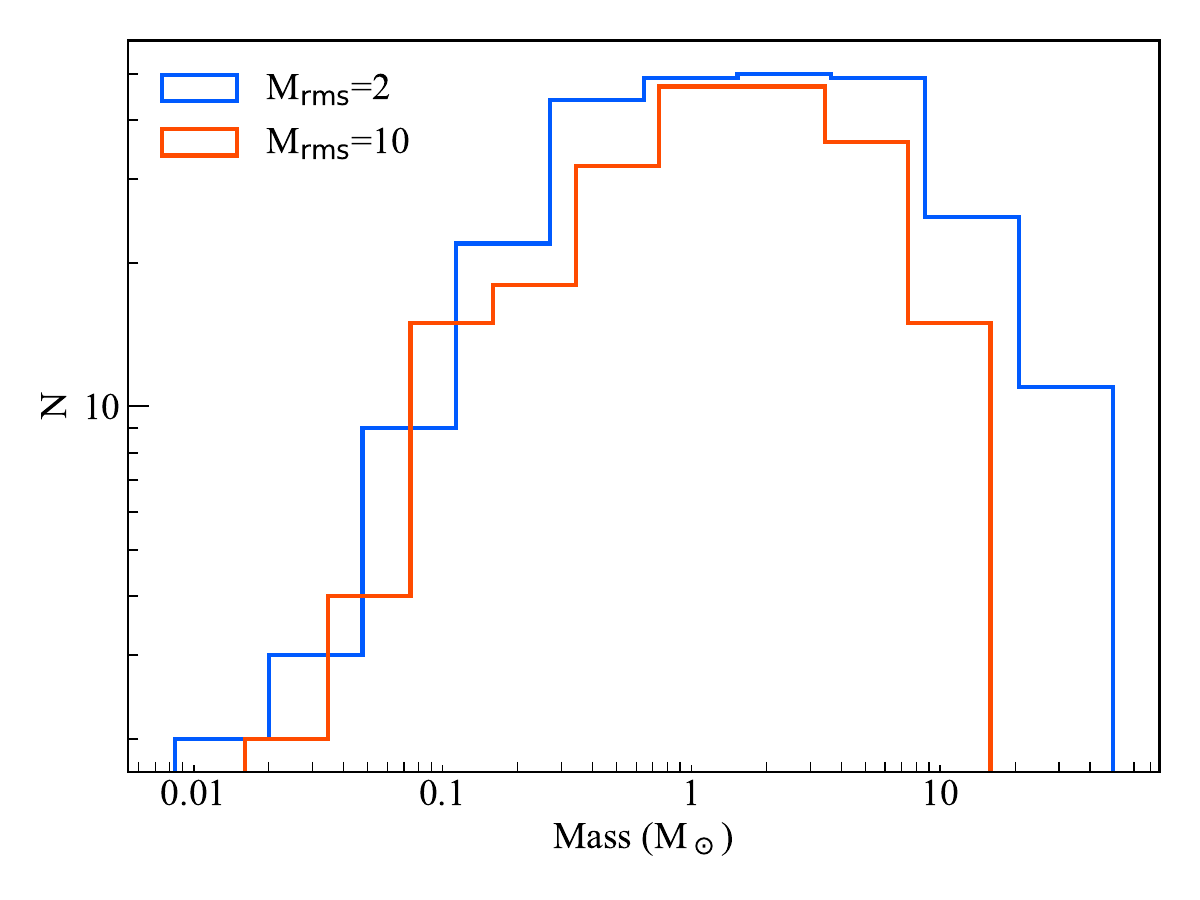}
    \caption{Mass distribution of protostars at the end of the simulation for different turbulent strengths. The blue line represents the mass function for $\mathcal{M}_{\rm rms} = 2$ at 2.66 Myr with a total of 264 protostars. The red line shows the mass function for $\mathcal{M}_{\rm rms} = 10$ at 1.72 Myr with a total of 217 protostars.}
    \label{imf}
\end{figure}

We conducted clump-scale hydrodynamic simulations with the sink particle method and heating and cooling terms. Figure \ref{imf} shows the mass distribution of protostars one million years after the first star formation under different turbulent strengths. Due to differences in the basic equations, it is challenging to directly compare the mass distribution obtained in this study with the stellar initial mass functions (IMF) derived from isothermal ideal MHD turbulence simulations \citep[e.g., ][]{padoan2002, padoan2014, federrath2017, haugbolle2018, pelkonen2021}.

The model of $\mathcal{M}_{\rm rms} = 10$ without converging flows in \cite{matsumoto2015} sets up similar initial conditions to our study, except for not considering heating and cooling terms. The mass distribution of sink particles obtained in that model largely agrees with the mass distribution from our results in $\mathcal{M}_{\rm rms} = 10$. Additionally, in \cite{matsumoto2015}, figure 12 suggests that, despite differences in turbulence strengths ($\mathcal{M}_{\rm rms} = 3, 10, 30$), the mass distribution consistently peaks around $\sim 1 \, \msun$, which is consistent with our results.

Similarly, while our initial conditions are similar to those in \cite{pelkonen2021}, differences such as magnetic fields and sink particle criteria exist. In Fig. \ref{imf}, compared to \cite{pelkonen2021}, the mass distribution for Mach 10 shows a shallower slope on the high-mass end and the mass distribution peaks at higher masses. This difference likely arises from variations in base grid sizes and the conditions for sink particles, as suggested by \cite{pelkonen2021}.

\section{Geometric Analysis of Shape in Star-Forming Cores} \label{asec:shape}
To analyze the shape of identified star-forming cores at various densities, we estimate the axis ratio in the principal axis coordinates. This ratio is determined by calculating the moment of inertia tensor {\bf I}, utilizing the physical properties such as the mass $m_i$ and position $r_i$ of each cell, where $r_i$ is measured as the distance from the position of the protostar, within the identified star-forming cores, as detailed in equations \eqref{momtensor} and \eqref{momtensor1}.
\begin{subequations}
\begin{align}
    I_{xx} &= \sum_{i} m_i (r_{iy}^2 + r_{iz}^2). \\
    I_{yy} &= \sum_{i} m_i (r_{ix}^2 + r_{iz}^2). \\
    I_{zz} &= \sum_{i} m_i (r_{ix}^2 + r_{iy}^2). \\
    I_{xy} &= I_{yx} = -\sum_{i} m_i r_{ix} r_{iy}. \\
    I_{xz} &= I_{zx} = -\sum_{i} m_i r_{ix} r_{iz}. \\
    I_{yz} &= I_{zy} = -\sum_{i} m_i r_{iy} r_{iz}.
\end{align}
    \label{momtensor}
\end{subequations}
\begin{equation}
    \mathbf{I} = \begin{pmatrix}
    I_{xx} & I_{xy} & I_{xz} \\
    I_{yx} & I_{yy} & I_{yz} \\
    I_{zx} & I_{zy} & I_{zz}
    \label{momtensor1}
\end{pmatrix} 
\end{equation}
The eigenvalues \(\lambda_i\) of the moment of inertia tensor were calculated, and the axis ratio \(\gamma_{\rm{core}}\) was determined by taking the ratio of the maximum eigenvalue \(\lambda_{\rm max}\) to the minimum eigenvalue \(\lambda_{\rm min}\), as shown in equations \eqref{eigenvector}.
\begin{equation}
     \mathbf{I} \mathbf{v}_i = \lambda_i \mathbf{v}_i 
     \label{eigenvector}
\end{equation}
The detailed results are presented in Section \ref{subsec:gasmass}.

\bibliography{rf}{}

\begin{thebibliography}{}
\expandafter\ifx\csname natexlab\endcsname\relax\def\natexlab#1{#1}\fi
\providecommand{\url}[1]{\href{#1}{#1}}
\providecommand{\dodoi}[1]{doi:~\href{http://doi.org/#1}{\nolinkurl{#1}}}
\providecommand{\doeprint}[1]{\href{http://ascl.net/#1}{\nolinkurl{http://ascl.net/#1}}}
\providecommand{\doarXiv}[1]{\href{https://arxiv.org/abs/#1}{\nolinkurl{https://arxiv.org/abs/#1}}}

\bibitem[{{Alves} {et~al.}(2001){Alves}, {Lada}, \& {Lada}}]{alves2001}
{Alves}, J.~F., {Lada}, C.~J., \& {Lada}, E.~A. 2001, Nature, 409, 159

\bibitem[{{Andr{\'e}} {et~al.}(2014){Andr{\'e}}, {Di Francesco}, {Ward-Thompson}, {Inutsuka}, {Pudritz}, \& {Pineda}}]{andre2014}
{Andr{\'e}}, P., {Di Francesco}, J., {Ward-Thompson}, D., {et~al.} 2014, in Protostars and Planets VI, ed. H.~{Beuther}, R.~S. {Klessen}, C.~P. {Dullemond}, \& T.~{Henning}, 27, \dodoi{10.2458/azu_uapress_9780816531240-ch002}

\bibitem[{{Andr{\'e}} {et~al.}(2010){Andr{\'e}}, {Men'shchikov}, {Bontemps}, {K{\"o}nyves}, {Motte}, {Schneider}, {Didelon}, {Minier}, {Saraceno}, {Ward-Thompson}, {di Francesco}, {White}, {Molinari}, {Testi}, {Abergel}, {Griffin}, {Henning}, {Royer}, {Mer{\'\i}n}, {Vavrek}, {Attard}, {Arzoumanian}, {Wilson}, {Ade}, {Aussel}, {Baluteau}, {Benedettini}, {Bernard}, {Blommaert}, {Cambr{\'e}sy}, {Cox}, {di Giorgio}, {Hargrave}, {Hennemann}, {Huang}, {Kirk}, {Krause}, {Launhardt}, {Leeks}, {Le Pennec}, {Li}, {Martin}, {Maury}, {Olofsson}, {Omont}, {Peretto}, {Pezzuto}, {Prusti}, {Roussel}, {Russeil}, {Sauvage}, {Sibthorpe}, {Sicilia-Aguilar}, {Spinoglio}, {Waelkens}, {Woodcraft}, \& {Zavagno}}]{andre2010}
{Andr{\'e}}, P., {Men'shchikov}, A., {Bontemps}, S., {et~al.} 2010, \aap, 518, L102, \dodoi{10.1051/0004-6361/201014666}

\bibitem[{{Bastian} {et~al.}(2010){Bastian}, {Meyer}, {Greissl}, \& {Seth}}]{bastian2010}
{Bastian}, N., {Meyer}, M., {Greissl}, J., \& {Seth}, A. 2010, {Testing IMF universality through the direct detection of low mass stars in starburst galaxies}, NOAO Proposal ID 2010B-0189

\bibitem[{{Battersby} {et~al.}(2010){Battersby}, {Bally}, {Jackson}, {Ginsburg}, {Shirley}, {Schlingman}, \& {Glenn}}]{Battersby2010}
{Battersby}, C., {Bally}, J., {Jackson}, J.~M., {et~al.} 2010, \apj, 721, 222, \dodoi{10.1088/0004-637X/721/1/222}

\bibitem[{{Bonnell} \& {Bate}(2006)}]{bonnell2006}
{Bonnell}, I.~A., \& {Bate}, M.~R. 2006, MNRASras, 370, 488, \dodoi{10.1111/j.1365-2966.2006.10495.x}

\bibitem[{{Chabrier}(2003)}]{chabrier2003}
{Chabrier}, G. 2003, \pasp, 115, 763, \dodoi{10.1086/376392}

\bibitem[{{Chen} {et~al.}(2019){Chen}, {Pineda}, {Goodman}, {Burkert}, {Offner}, {Friesen}, {Myers}, {Alves}, {Arce}, {Caselli}, {Chac{\'o}n-Tanarro}, {Chen}, {Di Francesco}, {Ginsburg}, {Keown}, {Kirk}, {Martin}, {Matzner}, {Punanova}, {Redaelli}, {Rosolowsky}, {Scibelli}, {Seo}, {Shirley}, {Singh}, \& {GAS Collaboration}}]{chen2019}
{Chen}, H. H.-H., {Pineda}, J.~E., {Goodman}, A.~A., {et~al.} 2019, \apj, 877, 93, \dodoi{10.3847/1538-4357/ab1a40}

\bibitem[{{Collins} {et~al.}(2023){Collins}, {Le}, \& {Jimenez Vela}}]{david2023}
{Collins}, D.~C., {Le}, D., \& {Jimenez Vela}, L.~L. 2023, \mnras, 520, 4194, \dodoi{10.1093/mnras/stac2834}

\bibitem[{{Collins} {et~al.}(2024){Collins}, {Le}, \& {Jimenez Vela}}]{david2024}
{Collins}, D.~C., {Le}, D.~K., \& {Jimenez Vela}, L.~L. 2024, \mnras, 532, 681, \dodoi{10.1093/mnras/stae1493}

\bibitem[{{Falgarone} {et~al.}(1992){Falgarone}, {Puget}, \& {Perault}}]{Falgarone1992}
{Falgarone}, E., {Puget}, J.~L., \& {Perault}, M. 1992, \aap, 257, 715

\bibitem[{{Federrath} {et~al.}(2017){Federrath}, {Krumholz}, \& {Hopkins}}]{federrath2017}
{Federrath}, C., {Krumholz}, M., \& {Hopkins}, P.~F. 2017, in Journal of Physics Conference Series, Vol. 837, Journal of Physics Conference Series (IOP), 012007, \dodoi{10.1088/1742-6596/837/1/012007}

\bibitem[{{Federrath} {et~al.}(2010){Federrath}, {Roman-Duval}, {Klessen}, {Schmidt}, \& {Mac Low}}]{federrath2010}
{Federrath}, C., {Roman-Duval}, J., {Klessen}, R.~S., {Schmidt}, W., \& {Mac Low}, M.~M. 2010, \aap, 512, A81, \dodoi{10.1051/0004-6361/200912437}

\bibitem[{{Fukushima} \& {Yajima}(2021)}]{fukushima2021}
{Fukushima}, H., \& {Yajima}, H. 2021, \mnras, 506, 5512, \dodoi{10.1093MNRAS/stab2099}

\bibitem[{{Fukushima} {et~al.}(2020){Fukushima}, {Yajima}, {Sugimura}, {Hosokawa}, {Omukai}, \& {Matsumoto}}]{fukushima2020}
{Fukushima}, H., {Yajima}, H., {Sugimura}, K., {et~al.} 2020, \mnras, 497, 3830, \dodoi{10.1093/mnras/staa2062}

\bibitem[{{Gong} \& {Ostriker}(2015)}]{gong2015}
{Gong}, M., \& {Ostriker}, E.~C. 2015, \apj, 806, 31, \dodoi{10.1088/0004-637X/806/1/31}

\bibitem[{{Goodman} {et~al.}(2009){Goodman}, {Rosolowsky}, {Borkin}, {Foster}, {Halle}, {Kauffmann}, \& {Pineda}}]{goodman2009}
{Goodman}, A.~A., {Rosolowsky}, E.~W., {Borkin}, M.~A., {et~al.} 2009, Nature, 457, 63, \dodoi{10.1038/nature07609}

\bibitem[{{Haugb{\o}lle} {et~al.}(2018){Haugb{\o}lle}, {Padoan}, \& {Nordlund}}]{haugbolle2018}
{Haugb{\o}lle}, T., {Padoan}, P., \& {Nordlund}, {\r{A}}. 2018, \apj, 854, 35, \dodoi{10.3847/1538-4357/aaa432}

\bibitem[{{Hirano} {et~al.}(2020){Hirano}, {Tsukamoto}, {Basu}, \& {Machida}}]{hirano2020}
{Hirano}, S., {Tsukamoto}, Y., {Basu}, S., \& {Machida}, M.~N. 2020, \apj, 898, 118, \dodoi{10.3847/1538-4357/ab9f9d}

\bibitem[{{Inutsuka} {et~al.}(2015){Inutsuka}, {Inoue}, {Iwasaki}, \& {Hosokawa}}]{inutsuka2015}
{Inutsuka}, S.-i., {Inoue}, T., {Iwasaki}, K., \& {Hosokawa}, T. 2015, \aap, 580, A49, \dodoi{10.1051/0004-6361/201425584}

\bibitem[{{Kandori} {et~al.}(2005){Kandori}, {Nakajima}, {Tamura}, {Tatematsu}, {Aikawa}, {Naoi}, {Sugitani}, {Nakaya}, {Nagayama}, {Nagata}, {Kurita}, {Kato}, {Nagashima}, \& {Sato}}]{kandori2005}
{Kandori}, R., {Nakajima}, Y., {Tamura}, M., {et~al.} 2005, AJ, 130, 2166, \dodoi{10.1086/444619}

\bibitem[{{Kerr} {et~al.}(2019){Kerr}, {Kirk}, {Di Francesco}, {Keown}, {Chen}, {Rosolowsky}, {Offner}, {Friesen}, {Pineda}, {Shirley}, {Redaelli}, {Caselli}, {Punanova}, {Seo}, {Alves}, {Chac{\'o}n-Tanarro}, \& {Chen}}]{kerr2019}
{Kerr}, R., {Kirk}, H., {Di Francesco}, J., {et~al.} 2019, \apj, 874, 147, \dodoi{10.3847/1538-4357/ab0c08}

\bibitem[{{Kirkpatrick} {et~al.}(2024){Kirkpatrick}, {Marocco}, {Gelino}, {Raghu}, {Faherty}, {Bardalez Gagliuffi}, {Schurr}, {Apps}, {Schneider}, {Meisner}, {Kuchner}, {Caselden}, {Smart}, {Casewell}, {Raddi}, {Kesseli}, {Stevnbak Andersen}, {Antonini}, {Beaulieu}, {Bickle}, {Bilsing}, {Chieng}, {Colin}, {Deen}, {Dereveanco}, {Doll}, {Durantini Luca}, {Frazer}, {Gantier}, {Gramaize}, {Grant}, {Hamlet}, {Higashimura}, {Hyogo}, {Ja{\l}owiczor}, {Jonkeren}, {Kabatnik}, {Kiwy}, {Martin}, {Michaels}, {Pendrill}, {Pessanha Machado}, {Pumphrey}, {Rothermich}, {Russwurm}, {Sainio}, {Sanchez}, {Sapelkin-Tambling}, {Sch{\"u}mann}, {Selg-Mann}, {Singh}, {Stenner}, {Sun}, {Tanner}, {Th{\'e}venot}, {Ventura}, {Voloshin}, {Walla}, {W{\k{e}}dracki}, {Adorno}, {Aganze}, {Allers}, {Brooks}, {Burgasser}, {Calamari}, {Connor}, {Costa}, {Eisenhardt}, {Gagn{\'e}}, {Gerasimov}, {Gonzales}, {Hsu}, {Kiman}, {Li}, {Low}, {Mamajek}, {Pantoja}, {Popinchalk}, {Rees}, {Stern}, {Su{\'a}rez}, {Theissen}, {Tsai}, {Vos}, {Zurek}, \& {The
  Backyard Worlds: Planet 9 Collaboration}}]{Kirkpatrick2024}
{Kirkpatrick}, J.~D., {Marocco}, F., {Gelino}, C.~R., {et~al.} 2024, \apjs, 271, 55, \dodoi{10.3847/1538-4365/ad24e2}

\bibitem[{{Koga} {et~al.}(2022){Koga}, {Kawasaki}, \& {Machida}}]{koga2022}
{Koga}, S., {Kawasaki}, Y., \& {Machida}, M.~N. 2022, \mnras, 515, 6073, \dodoi{10.1093MNRAS/stac2115}

\bibitem[{{Kong} {et~al.}(2015){Kong}, {Caselli}, {Tan}, {Wakelam}, \& {Sipil{\"a}}}]{kong2015}
{Kong}, S., {Caselli}, P., {Tan}, J.~C., {Wakelam}, V., \& {Sipil{\"a}}, O. 2015, \apj, 804, 98, \dodoi{10.1088/0004-637X/804/2/98}

\bibitem[{{K{\"o}nyves} {et~al.}(2010){K{\"o}nyves}, {Andr{\'e}}, {Men'shchikov}, {Schneider}, {Arzoumanian}, {Bontemps}, {Attard}, {Motte}, {Didelon}, {Maury}, {Abergel}, {Ali}, {Baluteau}, {Bernard}, {Cambr{\'e}sy}, {Cox}, {di Francesco}, {di Giorgio}, {Griffin}, {Hargrave}, {Huang}, {Kirk}, {Li}, {Martin}, {Minier}, {Molinari}, {Olofsson}, {Pezzuto}, {Russeil}, {Roussel}, {Saraceno}, {Sauvage}, {Sibthorpe}, {Spinoglio}, {Testi}, {Ward-Thompson}, {White}, {Wilson}, {Woodcraft}, \& {Zavagno}}]{konyves2010}
{K{\"o}nyves}, V., {Andr{\'e}}, P., {Men'shchikov}, A., {et~al.} 2010, \aap, 518, L106, \dodoi{10.1051/0004-6361/201014689}

\bibitem[{{K{\"o}nyves} {et~al.}(2015){K{\"o}nyves}, {Andr{\'e}}, {Men'shchikov}, {Palmeirim}, {Arzoumanian}, {Schneider}, {Roy}, {Didelon}, {Maury}, {Shimajiri}, {Di Francesco}, {Bontemps}, {Peretto}, {Benedettini}, {Bernard}, {Elia}, {Griffin}, {Hill}, {Kirk}, {Ladjelate}, {Marsh}, {Martin}, {Motte}, {Nguy{\^e}n Luong}, {Pezzuto}, {Roussel}, {Rygl}, {Sadavoy}, {Schisano}, {Spinoglio}, {Ward-Thompson}, \& {White}}]{konyves2015}
---. 2015, \aap, 584, A91, \dodoi{10.1051/0004-6361/201525861}

\bibitem[{{K{\"o}nyves} {et~al.}(2020){K{\"o}nyves}, {Andr{\'e}}, {Arzoumanian}, {Schneider}, {Men'shchikov}, {Bontemps}, {Ladjelate}, {Didelon}, {Pezzuto}, {Benedettini}, {Bracco}, {Di Francesco}, {Goodwin}, {Rygl}, {Shimajiri}, {Spinoglio}, {Ward-Thompson}, \& {White}}]{konyves2020}
{K{\"o}nyves}, V., {Andr{\'e}}, P., {Arzoumanian}, D., {et~al.} 2020, \aap, 635, A34, \dodoi{10.1051/0004-6361/201834753}

\bibitem[{{Kroupa}(2001)}]{kroupa2001}
{Kroupa}, P. 2001, \mnras, 322, 231, \dodoi{10.1046/j.1365-8711.2001.04022.x}

\bibitem[{{Kumar} {et~al.}(2020){Kumar}, {Palmeirim}, {Arzoumanian}, \& {Inutsuka}}]{kumar2020}
{Kumar}, M.~S.~N., {Palmeirim}, P., {Arzoumanian}, D., \& {Inutsuka}, S.~I. 2020, \aap, 642, A87, \dodoi{10.1051/0004-6361/202038232}

\bibitem[{{Ladjelate} {et~al.}(2020){Ladjelate}, {Andr{\'e}}, {K{\"o}nyves}, {Ward-Thompson}, {Men'shchikov}, {Bracco}, {Palmeirim}, {Roy}, {Shimajiri}, {Kirk}, {Arzoumanian}, {Benedettini}, {Di Francesco}, {Fiorellino}, {Schneider}, {Pezzuto}, {Motte}, \& {Herschel Gould Belt Survey Team}}]{ladjelate2020}
{Ladjelate}, B., {Andr{\'e}}, P., {K{\"o}nyves}, V., {et~al.} 2020, \aap, 638, A74, \dodoi{10.1051/0004-6361/201936442}

\bibitem[{{Mac Low}(1999)}]{maclow1999}
{Mac Low}, M.-M. 1999, \apj, 524, 169, \dodoi{10.1086/307784}

\bibitem[{{Mac Low} \& {Klessen}(2004)}]{maclow2004}
{Mac Low}, M.-M., \& {Klessen}, R.~S. 2004, Reviews of Modern Physics, 76, 125, \dodoi{10.1103/RevModPhys.76.125}

\bibitem[{{Machida} \& {Hosokawa}(2013)}]{machida2013}
{Machida}, M.~N., \& {Hosokawa}, T. 2013, \mnras, 431, 1719, \dodoi{10.1093MNRAS/stt291}

\bibitem[{{Machida} \& {Matsumoto}(2012)}]{machida2012}
{Machida}, M.~N., \& {Matsumoto}, T. 2012, \mnras, 421, 588, \dodoi{10.1111/j.1365-2966.2011.20336.x}

\bibitem[{{Marsh} {et~al.}(2016){Marsh}, {Kirk}, {Andr{\'e}}, {Griffin}, {K{\"o}nyves}, {Palmeirim}, {Men'shchikov}, {Ward-Thompson}, {Benedettini}, {Bresnahan}, {di Francesco}, {Elia}, {Motte}, {Peretto}, {Pezzuto}, {Roy}, {Sadavoy}, {Schneider}, {Spinoglio}, \& {White}}]{marsh2016}
{Marsh}, K.~A., {Kirk}, J.~M., {Andr{\'e}}, P., {et~al.} 2016, \mnras, 459, 342, \dodoi{10.1093/mnras/stw301}

\bibitem[{{Matsumoto}(2007)}]{matsumoto2007}
{Matsumoto}, T. 2007, PASJ, 59, 905, \dodoi{10.1093/pasj/59.5.905}

\bibitem[{{Matsumoto} {et~al.}(2015){Matsumoto}, {Dobashi}, \& {Shimoikura}}]{matsumoto2015}
{Matsumoto}, T., {Dobashi}, K., \& {Shimoikura}, T. 2015, \apj, 801, 77, \dodoi{10.1088/0004-637X/801/2/77}

\bibitem[{{McKee} \& {Ostriker}(1977)}]{mckee1977}
{McKee}, C.~F., \& {Ostriker}, J.~P. 1977, \apj, 218, 148, \dodoi{10.1086/155667}

\bibitem[{{McKee} \& {Tan}(2003)}]{mckee2003}
{McKee}, C.~F., \& {Tan}, J.~C. 2003, ApJj, 585, 850, \dodoi{10.1086/346149}

\bibitem[{{Moseley} {et~al.}(2023){Moseley}, {Teyssier}, \& {Draine}}]{moseley2023}
{Moseley}, E.~R., {Teyssier}, R., \& {Draine}, B.~T. 2023, \mnras, 518, 2825, \dodoi{10.1093MNRAS/stac3231}

\bibitem[{{Motte} \& {Andr{\'e}}(2001)}]{motte2001}
{Motte}, F., \& {Andr{\'e}}, P. 2001, \aap, 365, 440, \dodoi{10.1051/0004-6361:20000072}

\bibitem[{{Motte} {et~al.}(1998){Motte}, {Andre}, \& {Neri}}]{motte1998}
{Motte}, F., {Andre}, P., \& {Neri}, R. 1998, \aap, 336, 150

\bibitem[{{Nozaki} \& {Machida}(2023)}]{nozaki2023}
{Nozaki}, S., \& {Machida}, M.~N. 2023, \mnras, 519, 5017, \dodoi{10.1093/mnras/stac3819}

\bibitem[{{Nutter} \& {Ward-Thompson}(2007)}]{nutter2007}
{Nutter}, D., \& {Ward-Thompson}, D. 2007, \mnras, 374, 1413, \dodoi{10.1111/j.1365-2966.2006.11246.x}

\bibitem[{{Offner} {et~al.}(2022){Offner}, {Taylor}, {Markey}, {Chen}, {Pineda}, {Goodman}, {Burkert}, {Ginsburg}, \& {Choudhury}}]{offner2022}
{Offner}, S. S.~R., {Taylor}, J., {Markey}, C., {et~al.} 2022, \mnras, 517, 885, \dodoi{10.1093MNRAS/stac2734}

\bibitem[{{Onishi} {et~al.}(2002){Onishi}, {Mizuno}, {Kawamura}, {Tachihara}, \& {Fukui}}]{onishi2002}
{Onishi}, T., {Mizuno}, A., {Kawamura}, A., {Tachihara}, K., \& {Fukui}, Y. 2002, \apj, 575, 950, \dodoi{10.1086/341347}

\bibitem[{{Padoan} {et~al.}(2014){Padoan}, {Haugb{\o}lle}, \& {Nordlund}}]{padoan2014}
{Padoan}, P., {Haugb{\o}lle}, T., \& {Nordlund}, {\r{A}}. 2014, \apj, 797, 32, \dodoi{10.1088/0004-637X/797/1/32}

\bibitem[{{Padoan} \& {Nordlund}(2002)}]{padoan2002}
{Padoan}, P., \& {Nordlund}, {\r{A}}. 2002, \apj, 576, 870, \dodoi{10.1086/341790}

\bibitem[{{Padoan} {et~al.}(2007){Padoan}, {Nordlund}, {Kritsuk}, {Norman}, \& {Li}}]{padoan2007}
{Padoan}, P., {Nordlund}, {\r{A}}., {Kritsuk}, A.~G., {Norman}, M.~L., \& {Li}, P.~S. 2007, \apj, 661, 972, \dodoi{10.1086/516623}

\bibitem[{{Padoan} {et~al.}(2020){Padoan}, {Pan}, {Juvela}, {Haugb{\o}lle}, \& {Nordlund}}]{padoan2020}
{Padoan}, P., {Pan}, L., {Juvela}, M., {Haugb{\o}lle}, T., \& {Nordlund}, {\r{A}}. 2020, \apj, 900, 82, \dodoi{10.3847/1538-4357/abaa47}

\bibitem[{{Palmeirim} {et~al.}(2013){Palmeirim}, {Andr{\'e}}, {Kirk}, {Ward-Thompson}, {Arzoumanian}, {K{\"o}nyves}, {Didelon}, {Schneider}, {Benedettini}, {Bontemps}, {Di Francesco}, {Elia}, {Griffin}, {Hennemann}, {Hill}, {Martin}, {Men'shchikov}, {Molinari}, {Motte}, {Nguyen Luong}, {Nutter}, {Peretto}, {Pezzuto}, {Roy}, {Rygl}, {Spinoglio}, \& {White}}]{palmeirim2013}
{Palmeirim}, P., {Andr{\'e}}, P., {Kirk}, J., {et~al.} 2013, \aap, 550, A38, \dodoi{10.1051/0004-6361/201220500}

\bibitem[{{Pan} {et~al.}(2016){Pan}, {Padoan}, {Haugb{\o}lle}, \& {Nordlund}}]{pan2016}
{Pan}, L., {Padoan}, P., {Haugb{\o}lle}, T., \& {Nordlund}, {\r{A}}. 2016, \apj, 825, 30, \dodoi{10.3847/0004-637X/825/1/30}

\bibitem[{{Pelkonen} {et~al.}(2021){Pelkonen}, {Padoan}, {Haugb{\o}lle}, \& {Nordlund}}]{pelkonen2021}
{Pelkonen}, V.~M., {Padoan}, P., {Haugb{\o}lle}, T., \& {Nordlund}, {\r{A}}. 2021, \mnras, 504, 1219, \dodoi{10.1093MNRAS/stab844}

\bibitem[{{Rathborne} {et~al.}(2007){Rathborne}, {Simon}, \& {Jackson}}]{Rathborne2007}
{Rathborne}, J.~M., {Simon}, R., \& {Jackson}, J.~M. 2007, \apj, 662, 1082, \dodoi{10.1086/513178}

\bibitem[{{Redaelli} {et~al.}(2022){Redaelli}, {Chac{\'o}n-Tanarro}, {Caselli}, {Tafalla}, {Pineda}, {Spezzano}, \& {Sipil{\"a}}}]{Redaelli2022}
{Redaelli}, E., {Chac{\'o}n-Tanarro}, A., {Caselli}, P., {et~al.} 2022, \apj, 941, 168, \dodoi{10.3847/1538-4357/ac9d8b}

\bibitem[{{Salpeter}(1955)}]{salpeter1955}
{Salpeter}, E.~E. 1955, \apj, 121, 161, \dodoi{10.1086/145971}

\bibitem[{{Sato} {et~al.}(2023){Sato}, {Takahashi}, {Ishii}, {Ho}, {Machida}, {Carpenter}, {A. Zapata}, {Teixeira}, \& {Suri}}]{sato2023}
{Sato}, A., {Takahashi}, S., {Ishii}, S., {et~al.} 2023, \apj, 944, 92, \dodoi{10.3847/1538-4357/aca7c9}

\bibitem[{{Seifried} \& {Walch}(2015)}]{seifried2015}
{Seifried}, D., \& {Walch}, S. 2015, \mnras, 452, 2410, \dodoi{10.1093/mnras/stv1458}

\bibitem[{{Shimajiri} {et~al.}(2019){Shimajiri}, {Andr{\'e}}, {Ntormousi}, {Men'shchikov}, {Arzoumanian}, \& {Palmeirim}}]{shimajiri2019}
{Shimajiri}, Y., {Andr{\'e}}, P., {Ntormousi}, E., {et~al.} 2019, \aap, 632, A83, \dodoi{10.1051/0004-6361/201935689}

\bibitem[{{Singh} {et~al.}(2021){Singh}, {Matzner}, {Friesen}, {Martin}, {Pineda}, {Rosolowsky}, {Alves}, {Chac{\'o}n-Tanarro}, {Chen}, {Chen}, {Choudhury}, {Di Francesco}, {Keown}, {Kirk}, {Punanova}, {Seo}, {Shirley}, {Ginsburg}, {Offner}, {Arce}, {Caselli}, {Goodman}, {Myers}, {Redaelli}, \& {GAS Collaboration}}]{singh2021}
{Singh}, A., {Matzner}, C.~D., {Friesen}, R.~K., {et~al.} 2021, \apj, 922, 87, \dodoi{10.3847/1538-4357/ac20d2}

\bibitem[{{Smith} {et~al.}(2009){Smith}, {Clarke}, \& {Harland}}]{smith2009}
{Smith}, D.~M., {Clarke}, G.~P., \& {Harland}, K. 2009, Environment and Planning A: Economy and Space, 41, 1251, \dodoi{10.1068/a4147}

\bibitem[{{Smullen} {et~al.}(2020){Smullen}, {Kratter}, {Offner}, {Lee}, \& {Chen}}]{smullen2020}
{Smullen}, R.~A., {Kratter}, K.~M., {Offner}, S. S.~R., {Lee}, A.~T., \& {Chen}, H. H.-H. 2020, \mnras, 497, 4517, \dodoi{10.1093/mnras/staa2253}

\bibitem[{{Takemura} {et~al.}(2021{\natexlab{a}}){Takemura}, {Nakamura}, {Kong}, {Arce}, {Carpenter}, {Ossenkopf-Okada}, {Klessen}, {Sanhueza}, {Shimajiri}, {Tsukagoshi}, {Kawabe}, {Ishii}, {Dobashi}, {Shimoikura}, {Goldsmith}, {S{\'a}nchez-Monge}, {Kauffmann}, {Pillai}, {Padoan}, {Ginsberg}, {Smith}, {Bally}, {Mairs}, {Pineda}, {Lis}, {Burkhart}, {Schilke}, {Chen}, {Isella}, {Friesen}, {Goodman}, \& {Harper}}]{takemura2021a}
{Takemura}, H., {Nakamura}, F., {Kong}, S., {et~al.} 2021{\natexlab{a}}, \apjl, 910, L6, \dodoi{10.3847/2041-8213/abe7dd}

\bibitem[{{Takemura} {et~al.}(2021{\natexlab{b}}){Takemura}, {Nakamura}, {Ishii}, {Shimajiri}, {Sanhueza}, {Tsukagoshi}, {Kawabe}, {Hirota}, \& {Kataoka}}]{takemura2021b}
{Takemura}, H., {Nakamura}, F., {Ishii}, S., {et~al.} 2021{\natexlab{b}}, PASJ, 73, 487, \dodoi{10.1093/pasj/psab014}

\bibitem[{{Tatematsu} {et~al.}(1993){Tatematsu}, {Umemoto}, {Kameya}, {Hirano}, {Hasegawa}, {Hayashi}, {Iwata}, {Kaifu}, {Mikami}, {Murata}, {Nakano}, {Nakano}, {Ohashi}, {Sunada}, {Takaba}, \& {Yamamoto}}]{tatematsu1993}
{Tatematsu}, K., {Umemoto}, T., {Kameya}, O., {et~al.} 1993, \apj, 404, 643, \dodoi{10.1086/172318}

\bibitem[{{Tokuda} {et~al.}(2020){Tokuda}, {Fujishiro}, {Tachihara}, {Takashima}, {Fukui}, {Zahorecz}, {Saigo}, {Matsumoto}, {Tomida}, {Machida}, {Inutsuka}, {Andr{\'e}}, {Kawamura}, \& {Onishi}}]{tokuda2020}
{Tokuda}, K., {Fujishiro}, K., {Tachihara}, K., {et~al.} 2020, \apj, 899, 10, \dodoi{10.3847/1538-4357/ab9ca7}

\bibitem[{{Tokuda} {et~al.}(2022){Tokuda}, {Minami}, {Fukui}, {Inoue}, {Nishioka}, {Tsuge}, {Zahorecz}, {Sano}, {Konishi}, {Rosie Chen}, {Sewi{\l}o}, {Madden}, {Nayak}, {Saigo}, {Nishimura}, {Tanaka}, {Sawada}, {Indebetouw}, {Tachihara}, {Kawamura}, \& {Onishi}}]{tokuda2022}
{Tokuda}, K., {Minami}, T., {Fukui}, Y., {et~al.} 2022, \apj, 933, 20, \dodoi{10.3847/1538-4357/ac6b3c}

\bibitem[{{Tokuda} {et~al.}(2023){Tokuda}, {Harada}, {Tanaka}, {Inoue}, {Shimonishi}, {Zhang}, {Sewi{\l}o}, {Kunitoshi}, {Konishi}, {Fukui}, {Kawamura}, {Onishi}, \& {Machida}}]{tokuda2023}
{Tokuda}, K., {Harada}, N., {Tanaka}, K. E.~I., {et~al.} 2023, \apj, 955, 52, \dodoi{10.3847/1538-4357/acefb7}

\bibitem[{{Wang} {et~al.}(2010){Wang}, {Li}, {Abel}, \& {Nakamura}}]{wang2010}
{Wang}, P., {Li}, Z.-Y., {Abel}, T., \& {Nakamura}, F. 2010, \apj, 709, 27, \dodoi{10.1088/0004-637X/709/1/27}

\bibitem[{{Ward-Thompson} {et~al.}(2007){Ward-Thompson}, {Andr{\'e}}, {Crutcher}, {Johnstone}, {Onishi}, \& {Wilson}}]{ward-thompson2007}
{Ward-Thompson}, D., {Andr{\'e}}, P., {Crutcher}, R., {et~al.} 2007, in Protostars and Planets V, ed. B.~{Reipurth}, D.~{Jewitt}, \& K.~{Keil}, 33, \dodoi{10.48550/arXiv.astro-ph/0603474}

\bibitem[{{Williams} {et~al.}(1994){Williams}, {de Geus}, \& {Blitz}}]{williams1994}
{Williams}, J.~P., {de Geus}, E.~J., \& {Blitz}, L. 1994, \apj, 428, 693, \dodoi{10.1086/174279}

\end{thebibliography}
\bibliographystyle{aasjournal}

\end{document}